\documentclass[a4paper,12pt]{article}
\pdfoutput=1

\usepackage{amsmath,amssymb,bbm,multirow,hyperref}
\usepackage{fullpage}

\DeclareMathOperator{\ud}{d}

\newtheorem{theorem}{Theorem}

\begin{document}

\title{Supersymmetry and R-symmetries\\
       in Wess-Zumino models:\\
       properties and model dataset construction}
\author{Zheng Sun\\
        \normalsize\textit{College of Physics, Sichuan University,}\\
        \normalsize\textit{29 Wangjiang Road, Chengdu 610064, P.~R.~China}\\
        \normalsize\textit{E-mail:} \texttt{sun\_ctp@scu.edu.cn}
       }
\date{}
\maketitle

\begin{abstract}
The Nelson-Seiberg theorem and its extensions relate supersymmetry breaking and R-symmetries in Wess-Zumino models.  But their applicability may be limited by previously found non-generic counterexamples.  Constructing a dataset of R-symmetric Wess-Zumino models is useful for studying the occurrence of such counterexamples as well as other purposes.  This work gives a pedagogical review on the basics of supersymmetry in (3+1)-dimensions, Wess-Zumino models and their supergravity extensions, the Nelson-Seiberg theorem and its extensions.  We present a preliminary construction of the dataset of R-symmetric Wess-Zumino models with up to $5$ chiral fields.  Among $925$ models in total, $20$ of them with non-generic R-charges are counterexamples to both the Nelson-Seiberg theorem and its extensions.  Thus the dataset gives an estimation of the accuracy of the field counting method based on these theorems.  More constructions and applications of the dataset are expected in future work.
\end{abstract}

\section{Introduction}

The Standard Model (SM) has been well tested by current particle physics experiments and astrophysics observations.  The LHC experiment completed its last piece by the discovery of the Higgs boson in 2012~\cite{ATLAS:2012yve, CMS:2012qbp}, and since then verified SM predictions at high precision~\cite{LHCHiggsCrossSectionWorkingGroup:2013rie, Buttazzo:2013uya, ATLAS:2016neq, ParticleDataGroup:2018ovx, ParticleDataGroup:2020ssz}.  But there are still motivations to explore new theories beyond the Standard Model (BSM), for instance, the gauge hierarchy problem, the dark matter candidates, and unification of all interactions including gravity.  Supersymmetry (SUSY)~\cite{Nilles:1983ge, Haber:1984rc, Martin:1997ns, Quevedo:2010ui, Wess:1992cp, Bailin:1994qt, Baer:2006rs, Terning:2006bq, Dine:2007zp} is one of many proposals to address these problems.  In the SUSY extension of SM, gauge coupling unification can be realized, and the lightest supersymmetric particle (sparticle) becomes a dark matter candidate.  In string theory~\cite{Becker:2006dvp, Polchinski:1998rq, Polchinski:1998rr, Green:2012oqa, Green:2012pqa, Ibanez:2012zz, Blumenhagen:2013fgp} which is a candidate of unification theory including quantum gravity, SUSY is an important ingredient to eliminate unwanted tachyon states via the Gliozzi-Scherk-Olive (GSO) projection~\cite{Gliozzi:1976qd}.  These are examples that SUSY found its applications in both phenomenology and formal theory studies.

To build a realistic phenomenology model, spontaneous SUSY breaking has to occur in a hidden sector~\cite{Intriligator:2007cp}, be mediated to the visible sector, and give soft masses~\cite{Dimopoulos:1981zb} to sparticles which are beyond current experimental search limit.  Wess-Zumino models~\cite{Wess:1973kz, Wess:1974jb} serve as low-energy effective descriptions of dynamical SUSY breaking in the hidden sector~\cite{Intriligator:2007cp, Witten:1981nf, Affleck:1983mk, Affleck:1983vc, Affleck:1984xz, Dine:1993yw, Dine:1994vc, Dine:1995ag, Shadmi:1999jy, Intriligator:2006dd, Dine:2010cv}.  Their supergravity (SUGRA) extensions~\cite{Nilles:1983ge, Wess:1992cp, Bailin:1994qt, VanNieuwenhuizen:1981ae, Kallosh:2000ve, Freedman:2012zz} also serve as low-energy effective descriptions of Type IIB flux compactification in string phenomenology~\cite{Grana:2005jc, Douglas:2006es, Blumenhagen:2006ci, Denef:2007pq, Samtleben:2008pe, Lust:2009kp}.  SUSY breaking in Wess -Zumino models is related to R-symmetries by the Nelson-Seiberg theorem and its extensions~\cite{Nelson:1993nf, Sun:2011fq, Kang:2012fn, Li:2020wdk}.  Both SUSY and SUSY breaking vacua can be obtained by properly arranging fields with different R-charges.  These theorems lead to a field-counting method which is used to predict SUSY and SUSY breaking vacua without solving SUSY equations, and allows a quick survey or statistical study of different models~\cite{Dine:2005gz}.

The Nelson-Seiberg theorem and its extensions require genericity.  Apart from models with non-generic parameters which are unnatural, models with non-generic R-charges and generic parameters have been found to violate these theorems~\cite{Sun:2019bnd}.  These non-generic counterexamples may introduce non-neglectable error to the field-counting method.  Some counterexamples can be described as special arrangement of R-charges satisfying a sufficient condition~\cite{Amariti:2020lvx, Sun:2021svm, Li:2021ydn}, but new counterexamples not covered by the sufficient condition are also found~\cite{Brister:2022rrz}.  Here we present a preliminary construction of the dataset of R-symmetric Wess-Zumino models~\cite{Brister:2022vsz}.  Their vacuum solutions are examined and compared to the result of the field-counting method.  Thus the proportion of counterexamples in all models can be estimated.  The dataset may also give information for other purposes on request.

The rest part of this work is organized as following.  Section 2 gives a brief and pedagogical review of $\mathcal{N} = 1$ SUSY in (3+1)-dimensions, with SUSY breaking introduced at the end.  Section 3 reviews Wess-Zumino models, their SUGRA extensions, and the criteria for SUSY and SUSY breaking vacua in these models.  Section 4 reviews the Nelson-Seiberg theorem and its extensions, relating R-symmetries and the vacuum structure of certain models.  Section 5 presents the preliminary construction of a dataset of R-symmetric Wess-Zumino models, where the recorded data structure of models is specified, and a number of counterexamples are found in the dataset.  Section 6 gives some outlooks of issues and future expectations of this work.

\section{A brief review of SUSY in (3+1)-dimensions}

\subsection{The SUSY algebra}

Mathematically, SUSY is a Lie superalgebra extension of the space-time Poincar\'e symmetry.  For $\mathcal{N} = 1$ SUSY in (3+1)-dimensional Minkowski spacetime, in addition to the Poincar\'e algebra generators $P^\mu$ and $M^{\mu \nu}$, supercharge generators in the form of Weyl spinors $Q_\alpha$ and $\bar{Q}^{\dot{\alpha}}$ are introduced with anticommutation relations.  The full super-Poincar\'e algebra reads
\begin{align}
[P_\mu, P_\nu] &= 0,\\
[M_{\mu \nu}, P_\rho] &= i (\eta_{\nu \rho} P_\mu
                            - \eta_{\mu \rho} P_\nu),\\
[M_{\mu \nu}, M_{\rho \sigma}] &= i (\eta_{\nu \rho} M_{\mu \sigma}
                                     + \eta_{\mu \sigma} M_{\nu \rho}
                                     - \eta_{\nu \sigma} M_{\mu \rho}
                                     - \eta_{\mu \rho} M_{\nu \sigma}),\\
[Q_\alpha, P_\mu] &= [\bar{Q}_{\dot{\alpha}}, P_\mu]
                   = 0,\\
[Q_\alpha, M_{\mu \nu}]
&= i
   (\sigma_{\mu \nu})_{\alpha}^{\hphantom{\alpha} \beta}
   Q_\beta,\\
[\bar{Q}_{\dot{\alpha}}, M_{\mu \nu}]
&= - i
     \bar{Q}_{\dot{\beta}}
     (\bar{\sigma}_{\mu \nu})^{\dot{\beta}}_{\hphantom{\dot{\beta}} \dot{\alpha}},\\
\{ Q_\alpha, Q_\beta \} &= \{ \bar{Q}_{\dot{\alpha}}, \bar{Q}_{\dot{\beta}} \}
                         = 0,\\
\{ Q_\alpha, \bar{Q}_{\dot{\alpha}} \} &= 2 (\sigma^\mu)_{\alpha \dot{\alpha}} P_\mu,
\end{align}
where the Einstein summation convention for indices is adopted.  $\sigma^\mu$ is the 4-vector of Pauli matrices
\begin{equation}
(\sigma^\mu)_{\alpha \dot{\alpha}}
= (\mathbbm{1}, \vec \sigma)
= \left ( \begin{pmatrix} 1 & 0\\ 0 & 1 \end{pmatrix},
          \begin{pmatrix} 0 & 1\\ 1 & 0 \end{pmatrix},
          \begin{pmatrix} 0 & - i\\ i & 0 \end{pmatrix},
          \begin{pmatrix} 1 & 0\\ 0 & - 1 \end{pmatrix} \right ).
\end{equation}
Other related sigma matrices are given by
\begin{align}
(\bar{\sigma}^\mu)^{\dot{\alpha} \alpha} &= (\mathbbm{1}, - \vec \sigma),\\
(\sigma^{\mu \nu})_\alpha^{\hphantom{\alpha} \beta}
&= \frac{i}{4} ((\sigma^\mu)_{\alpha \dot{\alpha}}
                (\bar{\sigma}^\nu)^{\dot{\alpha} \beta}
                - (\sigma^\nu)_{\alpha \dot{\alpha}}
                  (\bar{\sigma}^\mu)^{\dot{\alpha} \beta}),\\
(\bar{\sigma}^{\mu \nu})^{\dot{\alpha}}_{\hphantom{\dot{\alpha}} \dot{\beta}}
&= \frac{i}{4} ((\bar{\sigma}^\mu)^{\dot{\alpha} \alpha}
                (\sigma^\nu)_{\alpha \dot{\beta}}
                - (\bar{\sigma}^\nu)^{\dot{\alpha} \alpha}
                  (\sigma^\mu)_{\alpha \dot{\beta}}).
\end{align}
We choose the convention of the spacetime metric
\begin{equation}
\eta_{\mu \nu} = \eta^{\mu \nu}
               = \operatorname{diag}(1, - 1, - 1, - 1)
\end{equation}
for raising and lowering spacetime indices, and analogously
\begin{equation}
\epsilon_{\alpha \beta} = \epsilon^{\alpha \beta}
                        = \epsilon_{\dot{\alpha} \dot{\beta}}
                        = \epsilon^{\dot{\alpha} \dot{\beta}}
                        = \epsilon
                        = \begin{pmatrix} 0 & 1\\ -1 & 0 \end{pmatrix} \label{eq:2-1-01}
\end{equation}
for raising and lowering spinor indices.

It is convenient to consider the superspace by introducing fermionic coordinates labeled in Grassmann numbers $\theta_\alpha$ and $\bar{\theta}^{\dot{\alpha}}$.  The super-Poincar\'e algebra represented as operators acting on functions of the superspace coordinates $(x^\mu, \theta_\alpha, \bar{\theta}^{\dot{\alpha}})$ is
\begin{align}
P_\mu &= i \partial_\mu,\\
Q_\alpha &= i \partial_\alpha
            - (\sigma^\mu)_{\alpha \dot{\alpha}}
              \bar{\theta}^{\dot{\alpha}}
              \partial_\mu,\\
\bar{Q}_{\dot{\alpha}} &= - i \bar{\partial}_{\dot{\alpha}}
                          + \theta^\alpha
                            (\sigma^\mu)_{\alpha \dot{\alpha}}
                            \partial_\mu,\\
M_{\mu \nu} &= i (x_\mu \partial_\nu
                  - x_\nu \partial_\mu
                  + \theta^\alpha
                    (\sigma_{\mu \nu})_\alpha^{\hphantom{\alpha} \beta}
                    \partial_\beta
                  + \bar{\theta}_{\dot{\alpha}}
                    (\bar{\sigma}_{\mu \nu})^{\dot{\alpha}}_{\hphantom{\dot{\alpha}} \dot{\beta}}
                    \bar{\partial}^{\dot{\beta}}).
\end{align}
A general element of the super-Poincar\'e group is given by
\begin{equation}
g = \exp (- i (\omega^{\mu \nu} M_{\mu \nu}
               + x^\mu P_\mu
               + \theta Q
               + \bar{\theta} \bar{Q})).
\end{equation}
From now on we use short notations of fermion bilinears
\begin{align}
\psi \chi
&= \psi^\alpha \chi_\alpha 
 = - \psi_\alpha \chi^\alpha
 = \chi \psi,\\
\bar{\psi} \bar{\chi}
&= \bar{\psi}_{\dot{\alpha}} \bar{\chi}^{\dot{\alpha}}
 = - \bar{\psi}^{\dot{\alpha}} \bar{\chi}_{\dot{\alpha}}
 = \bar{\chi} \bar{\psi},\\
\psi \sigma^\mu \bar{\chi}
&= \psi^\alpha
   (\sigma^\mu)_{\alpha \dot{\alpha}}
   \bar{\chi}^{\dot{\alpha}}
 = - \bar{\chi}_{\dot{\alpha}}
     (\bar{\sigma}^\mu)^{\dot{\alpha} \alpha}
     \psi_\alpha
 = - \bar{\chi} \bar{\sigma}^\mu \psi,\\
\psi \sigma^{\mu \nu} \chi
&= \psi^\alpha
   (\sigma^{\mu \nu})_\alpha^{\hphantom{\alpha} \beta}
   \chi_\beta
 = - \chi^\alpha
     (\sigma^{\mu \nu})_\alpha^{\hphantom{\alpha} \beta}
     \psi_\beta
 = - \chi \sigma^{\mu \nu} \psi,\\
\bar{\psi} \bar{\sigma}^{\mu \nu} \bar{\chi}
&= \bar{\psi}_{\dot{\alpha}}
   (\bar{\sigma}^{\mu \nu})^{\dot{\alpha}}_{\hphantom{\dot{\alpha}} \dot{\beta}}
   \bar{\chi}^{\dot{\beta}}
 = - \bar{\chi}_{\dot{\alpha}}
     (\bar{\sigma}^{\mu \nu})^{\dot{\alpha}}_{\hphantom{\dot{\alpha}} \dot{\beta}}
     \bar{\psi}^{\dot{\beta}}
 = - \bar{\chi} \bar{\sigma}^{\mu \nu} \bar{\psi}.
\end{align}

\subsection{Superfields}

Fundamental fields are classified according to their transformation properties under the super-Poincar\'e algebra, and identified as irreducible representations of the Lie superalgebra, or supermultiplets.  Using the language of superspace, supermultiplets are organized as superfields $S(x, \theta_\alpha, \bar{\theta}^{\dot{\alpha}})$, with the general expansion
\begin{equation}
\begin{split}
S &= a(x)
     + \theta \chi(x)
     + \bar{\theta} \bar{\chi}(x)
     + \theta \theta b(x)
     + \bar{\theta} \bar{\theta} c(x)\\
   &\quad 
     + \theta \sigma^\mu \bar{\theta} v_\mu(x)
     + \theta \theta \bar{\theta} \bar{\xi}(x)
     + \bar{\theta} \bar{\theta} \theta \xi(x)
     + \theta \theta \bar{\theta} \bar{\theta} d(x).
\end{split}
\end{equation}
To make derivatives of a superfield consistent with SUSY transformations, we introduce chiral and antichiral covariant derivatives
\begin{align}
\mathcal{D}_\alpha
&= \partial_\alpha
   - i (\sigma^\mu)_{\alpha \dot{\alpha}} \bar{\theta}^{\dot{\alpha}} \partial_\mu,\\
\bar{\mathcal{D}}_{\dot{\alpha}}
&= - \bar{\partial}_{\dot{\alpha}}
   + i \theta^\alpha (\sigma^\mu)_{\alpha \dot{\alpha}} \partial_\mu, \quad
\end{align}
which anticommute with supercharges.

The chiral superfield
\begin{equation}
\begin{split}
\Phi &= \phi(x)
        + \sqrt{2} \theta \psi(x)
        + \theta \theta F(x)\\
     &\quad
        + i \theta \sigma^\mu \bar{\theta} \partial_\mu \phi(x)
        - \frac{i}{\sqrt{2}} \theta \theta \partial_\mu \psi(x) \sigma^\mu \bar{\theta}
        - \frac{1}{4} \theta \theta \bar{\theta} \bar{\theta} \partial_\mu \partial^\mu \phi(x)
\end{split}
\end{equation}
satisfies $\bar{\mathcal{D}}_{\dot{\alpha}} \Phi = 0$.  It contains the scalar boson $\phi$, the left-handed fermion $\psi$, and the auxiliary field $F$.  Its conjugate $\bar{\Phi} = \Phi^*$, named the antichiral superfield and satisfying $D_\alpha \bar{\Phi}$, has similar field components except that the fermion component $\bar{\psi}$ becomes right-handed.  The vector superfield satisfying $V = V^*$ is usually related to a gauge group, and subject to the gauge transformation
\begin{equation}
V \to V + \frac{i}{2 g} (\Omega - \Omega^*)
\end{equation}
for an Abelian gauge group, and
\begin{equation}
e^{2 g V^a t^a} \to e^{-i \Omega^{a *} t^a} e^{2 g V^a t^a} e^{i \Omega^a t^a}.
\end{equation}
for a non-Abelian gauge group, where the gauge function $\Omega$ is a chiral superfield, and $t^a$ are generators of the non-Abelian gauge group with the commutator
\begin{equation}
[t^a, t^b] = i f_{a b c} t^c.
\end{equation}
Gauge transformations correspond to unphysical degrees of freedom in $V$.  It is necessary to eliminate such redundant degrees of freedom by gauge fixing.  Choosing the Wess-Zumino gauge, we have
\begin{equation}
V = \theta \sigma^\mu \bar{\theta} V_\mu(x)
    + i \theta \theta \bar{\theta} \bar{\lambda}(x)
    - i \bar{\theta} \bar{\theta} \theta \lambda(x)
    + \frac{1}{2} \theta \theta \bar{\theta} \bar{\theta} D(x),
\end{equation}
which contains the vector gauge boson $V_\mu$, the gauginos $\lambda$, $\bar{\lambda}$, and the auxiliary field $D$.  In analogy to the field strength in field theories, we have the field strength superfields
\begin{equation}
W_\alpha = - \frac{1}{4} \bar{\mathcal{D}} \bar{\mathcal{D}} \mathcal{D}_\alpha V
\end{equation}
for an Abelian gauge group, and
\begin{equation}
\begin{split}
W_\alpha = W^a_\alpha t^a
        &= - \frac{1}{8 g} \bar{\mathcal{D}} \bar{\mathcal{D}}
             (e^{- 2 g V^a t^a} \mathcal{D}_\alpha e^{2 g V^a t^a})\\
        &= - \frac{1}{4} \bar{\mathcal{D}} \bar{\mathcal{D}}
             (\mathcal{D}_\alpha V^a + i g f_{a b c} (\mathcal{D}_\alpha V^b) V^c)
             t^a
\end{split}
\end{equation}
for a non-Abelian gauge group, where $g$ is the gauge coupling constant.  A general superfield can always be decomposed to a chiral superfield, an antichiral superfield and a vector superfield, i.e.,
\begin{equation}
S = \Phi + \bar{\Phi} + V.
\end{equation}

\subsection{The SUSY Lagrangian}

To constructe the Lagrangian from superfields, we introduce the Berezin integral, or integration on Grassmann numbers $\theta_\alpha$ and $\bar{\theta}^{\dot{\alpha}}$, which satisfies
\begin{equation}
\int \! \ud \! \theta^\alpha
= \int \! \ud \! \bar{\theta}_{\dot{\alpha}}
= 0, \quad
\int \! \ud \! \theta^\alpha \, \theta_\alpha
= \int \! \ud \! \bar{\theta}_{\dot{\alpha}} \, \bar{\theta}^{\dot{\alpha}}
= 1, \label{eq:2-3-01}
\end{equation}
where the spinor indices are not summed in the second formula.  Using the short notations
\begin{equation}
\ud^2 \! \theta = \ud \! \theta^\alpha
                  \ud \! \theta_\alpha, \quad
\ud^2 \! \bar{\theta} = \ud \! \bar{\theta}_{\dot{\alpha}}
                        \ud \! \bar{\theta}^{\dot{\alpha}}, \quad
\ud^4 \! \theta = \ud^2 \! \theta
                  \ud^2 \! \bar{\theta}^2,
\end{equation}
we have the identities
\begin{equation}
\int \! \ud^2 \! \theta \, \theta \theta
= \int \! \ud^2 \! \bar{\theta} \, \bar{\theta} \bar{\theta}
= \int \! \ud^4 \! \theta \, \theta \theta \bar{\theta} \bar{\theta}
= 1.
\end{equation}

The most general Lagrangian density with SUSY, gauge invariance and renormalizability can be written as
\begin{equation}
\begin{split}
\mathcal{L}
&= \left ( \frac{1}{4} \int \! \ud^2 \! \theta \,
                       (W_I^{a \alpha} W^a_{I \alpha}
                        + W_J^{\alpha} W_{J \alpha})
           + \text{c.c.} \right )
   + \int \! \ud^4 \! \theta \,
     2 \xi_J V_{J}\\
&\quad
   + \int \! \ud^4 \! \theta \,
     \bar{\Phi}_i e^{2 g V_{(i)}} \Phi_i
   - \left ( \int \! \ud^2 \! \theta \,
             W(\Phi_i)
             + \text{c.c.} \right )\\
&= \left ( \frac{1}{4} [W_I^{a \alpha} W^a_{I \alpha}
                        + W_J^{\alpha} W_{J \alpha}]_{\theta \theta}
           + \text{c.c.} \right )
   + 2 \xi_J [V_{J}]_{\theta \theta \bar{\theta} \bar{\theta}}\\
&\quad
   + [\bar{\Phi}_i e^{2 g V_{(i)}} \Phi_i]_{\theta \theta \bar{\theta} \bar{\theta}}
   - ([W(\Phi_i)]_{\theta \theta} + \text{c.c.}). \label{eq:2-3-02}
\end{split}
\end{equation}
where $I$ labels non-Abelian gauge groups, $J$ labels Abelian $U(1)$ gauge groups, and $i$ labels chiral fields.  Other the than the usual gauge kinetic terms proportional to $ W_I^{a \alpha}  W^a_{I \alpha}$ or $W_J^{\alpha} W_{J \alpha}$, Fayet-Iliopoulos terms~\cite{Fayet:1974jb} with coefficients $\xi_J$ are introduced for $U(1)$ gauge groups.  The gauge coupling factor in the chiral kinetic term $\bar{\Phi}_i e^{2 g V_{(i)}} \Phi_i$ is defined as
\begin{equation}
g V_{(i)} = g_I V_I^a T_{(i) I}^a + g'_J q_{(i) J} V_J,
\end{equation}
where $g_I$ and $g_J$ are gauge coupling constants, $T_{(i) I}^a$ is the representation matrices of the non-Abelian gauge group generators $t_I^a$ acting on the represented space of $\Phi_i$, and $q_{(i) J}$ is the charge of $\Phi_i$ in the Abelian gauge group $U(1)_J$.  $W(\Phi_i)$, named the superpotential, is a holomorphic function of chiral fields.

\subsection{The MSSM Lagrangian}

The Lagrangian density \eqref{eq:2-3-02} covers all ingredients of the Minimal Supersymmetric Standard Model (MSSM)~\cite{Fayet:1974pd, Fayet:1976et, Fayet:1977yc}:  Quarks, leptons and their antiparticles are included in chiral and antichiral superfields, while gluons, weak bosons and the photon are included in vector superfields.  The Higgs field must extend to two $SU(2)$ doublets to cancel anomalies as well as to generate the needed Yukawa couplings.  These superfields and their representations in SM gauge groups are listed in Table \ref{tb:2-01}.  The superpotential is chosen to be
\begin{equation}
W = - (y_u)_{m n} Q_m \epsilon H_u \bar{U}_n
    + (y_d)_{m n} Q_m \epsilon H_d \bar{D}_n
    + (y_e)_{m n} L_m \epsilon H_d \bar{E}_n
    + \mu H_u \epsilon H_d,
\end{equation}
where $m$ and $n$ denote three families of quarks and leptons, and $\epsilon$ is the 2-dimensional antisymmetric matrix \eqref{eq:2-1-01} to contract two $SU(2)$ doublets to a singlet.  This superpotential leads to Yukawa couplings in SM\@.  It also gives quadratic terms of the Higgs fields but with wrong signs.  The correct Higgs potential terms must come from soft SUSY breaking terms which parameterize the high energy SUSY breaking effects at the electroweak scale.

\begin{table}
    \centering
    \renewcommand\arraystretch{1.3}
    \begin{tabular}{|c|c|c|c|c|c|c|}
        \hline
        \multicolumn{2}{|c|}{Superfields}                                & Bosons                         & Fermions                         & $SU(3)$            & $SU(2)$      & $U(1)$\\
        \hline
        \multirow{3}{6em}{\centering gauge bosons, gauginos} & $G$       & $g$                            & $\tilde{g}$                      & $\mathbf{8}$       & $\mathbf{1}$ & $\mathbf{0}$\\
        \cline{2-7}
                                                             & $W$       & $W^\pm, W^0$                   & $\tilde{W}^\pm, \tilde{W}^0$     & $\mathbf{1}$       & $\mathbf{3}$ & $\mathbf{0}$\\
        \cline{2-7}
                                                             & $B$       & $B^0$                          & $\tilde{B}^0$                    & $\mathbf{1}$       & $\mathbf{1}$ & $\mathbf{0}$\\
        \hline
        \multirow{3}{6em}{\centering squarks, quarks}        & $Q$       & $(\tilde{u}_L, \tilde{d}_L)$   & $(u_L, d_L)$                     & $\mathbf{3}$       & $\mathbf{2}$ & $\mathbf{1 / 6}$\\
        \cline{2-7}
                                                             & $\bar{U}$ & $\tilde{u}_R^*$                & $u_R^\dagger$                    & $\mathbf{\bar{3}}$ & $\mathbf{1}$ & $\mathbf{- 2 /3}$\\
        \cline{2-7}
                                                             & $\bar{D}$ & $\tilde{d}_R^*$                & $d_R^\dagger$                    & $\mathbf{\bar{3}}$ & $\mathbf{1}$ & $\mathbf{1 /3}$\\
        \hline
        \multirow{2}{6em}{\centering sleptons, leptons}      & $L$       & $(\tilde{\nu}_L, \tilde{e}_L)$ & $(\nu_L, e_L)$                   & $\mathbf{1}$       & $\mathbf{2}$ & $\mathbf{- 1 /2}$\\
        \cline{2-7}
                                                             & $\bar{E}$ & $\tilde{e}_R^*$                & $e_R^\dagger$                    & $\mathbf{1}$       & $\mathbf{1}$ & $\mathbf{1}$\\
        \hline
        \multirow{2}{6em}{\centering Higgs, Higgsinos}       & $H_u$     & $(H_u^+, H_u^0)$               & $(\tilde{H}_u^+, \tilde{H}_u^0)$ & $\mathbf{1}$       & $\mathbf{2}$ & $\mathbf{1 / 2}$\\
        \cline{2-7}
                                                             & $H_d$     & $(H_d^0, H_d^-)$               & $(\tilde{H}_d^0, \tilde{H}_d^-)$ & $\mathbf{1}$       & $\mathbf{2}$ & $\mathbf{- 1 / 2}$\\
        \hline
    \end{tabular}
    \caption{Superfields and their representations in SM gauge groups.  Right-handed fermions are written as their antiparticles, or charge conjugates, which are left-handed and included in chiral superfields.  The bars on $\bar{U}$, $\bar{D}$ or $\bar{E}$ are part of the names and should not be confused with the notation of antichiral superfields.}
    \label{tb:2-01}
\end{table}

\subsection{SUSY breaking}

Since none of the new particles in Table \ref{tb:2-01} has been discovered by experiments, SUSY must be spontaneously broken.  A SUSY breaking vacuum $| 0 \rangle$ is not invariant under SUSY transformations, i.e.,
\begin{equation}
Q_\alpha \lvert 0 \rangle \ne 0, \quad
\bar{Q}_{\dot{\alpha}} \lvert 0 \rangle \ne 0.
\end{equation}
The Hamiltonian operator $H$ can be expressed in term of SUSY generators as
\begin{equation}
H = P^0
  = \frac{1}{4}
    (\bar{\sigma}^0)^{\dot{\alpha} \alpha}
    \{ Q_\alpha, \bar{Q}_{\dot{\alpha}} \}
  = \frac{1}{4} (Q_1 \bar{Q}_{\dot{1}}
                 + \bar{Q}_{\dot{1}} Q_1
                 + Q_2 \bar{Q}_{\dot{2}}
                 + \bar{Q}_{\dot{2}} Q_2).
\end{equation}
The vacuum energy satisfies
\begin{equation}
\langle H \rangle = \langle 0 \rvert H \lvert 0 \rangle
                  = \frac{1}{4} (\lVert Q_1 \lvert 0 \rangle \rVert^2
                                 + \lVert \bar{Q}_{\dot{1}} \lvert 0 \rangle \rVert^2
                                 + \lVert Q_2 \lvert 0 \rangle \rVert^2
                                 + \lVert \bar{Q}_{\dot{2}} \lvert 0 \rangle \rVert^2)
                  \ge 0,
\end{equation}
and the equality is only satisfied by setting $Q_\alpha | 0 \rangle$ and $\bar{Q}_{\dot{\alpha}} | 0 \rangle$ to zero, which restores SUSY\@.  So $\langle H \rangle$ is the order parameter for SUSY breaking.  Neglecting spacetime dependency and fermion condensations at the vacuum, the Lagrangian density \eqref{eq:2-3-02} implies $\langle H \rangle = \langle V \rangle$, where $V$ is the scalar potential
\begin{equation}
V (\phi_i, \phi_i^*) = F_i^* F_i + \frac{1}{2} D_I^a D_I^a + \frac{1}{2} D_J D_J.
\end{equation}
Solving equations for auxiliary fields in the Lagrangian density \eqref{eq:2-3-02} gives their expectation values
\begin{align}
F_i^* &= \partial_i W
       = \frac{\partial W}{\partial \phi_i}, \quad
F_i = (\partial_i W)^*
    = \partial_{\bar{i}} W^*
    = \frac{\partial W^*}{\partial \phi_i^*},\\
D_I^a &= - g_I \phi_i^* T_{(i) I}^a \phi_i,\\
D_J &= - g_J q_{(i) J} \phi_i^* \phi_i - \xi_J,
\end{align}
all of which are evaluated at the expectation values of $\phi_i$.  Non-zero expectation values of either $F$'s or $D$'s break SUSY, corresponding to F-term or D-term SUSY breaking.  A realistic SUSY model has SUSY breaking in a hidden sector.  SUSY breaking effects are mediated to the visible sector, and give soft masses to sparticles which are beyond current experimental search limit.  Depending on the mediating interactions, there are scenarios of gravity mediation~\cite{Nilles:1983ge}, gauge mediation~\cite{Giudice:1998bp, Meade:2008wd} and anomaly mediation~\cite{Bagger:1999rd}, each of which has different phenomenological consequences.

\section{Wess-Zumino models and SUGRA extensions}

\subsection{Wess-Zumino models}

Both F-terms and D-terms can be the source for SUSY breaking in the hidden sector, and have been well studied for many years.  D-term SUSY breaking, usually involving Fayet-Iliopoulos terms~\cite{Fayet:1974jb}, are more difficult to give appropriate sparticle masses and to be consistent with quantum gravity~\cite{Komargodski:2009pc, Dienes:2009td, Komargodski:2010rb}.  F-term SUSY breaking models~\cite{Intriligator:2007cp}, also known as O'Raifeartaigh models~\cite{ORaifeartaigh:1975nky} or Wess-Zumino models~\cite{Wess:1973kz, Wess:1974jb}, have brighter phenomenological prospects.  They can also serve as low-energy effective descriptions of dynamical SUSY breaking models from Seiberg duality after integrating out strongly coupled gauge fields~\cite{Seiberg:1994bz, Intriligator:1994sm, Seiberg:1994pq, Intriligator:1996pu, Peskin:1997qi}.  In addition, it can be seen from their expressions that non-zero D-terms without Fayet-Iliopoulos terms must be accompanied with non-zero vacuum expectation values of $\phi_i$, which usually imply non-zero F-terms of a comparable scale.  For these reasons, we consider F-terms as the ultimate source of SUSY breaking in this work.

A Wess-Zumino model involves a set of chiral superfields $\{ \Phi_i \}$ and the holomorphic superpotential $W(\Phi_i)$.  The Lagrangian density for chiral superfields
\begin{equation}
\mathcal{L} = \int \! \ud^4 \! \theta \,
              \bar{\Phi}_i \Phi_i
              - \left ( \int \! \ud^2 \! \theta \,
                         W(\Phi_i) + \text{c.c.} \right ) \label{eq:3-1-01}
\end{equation}
leads to the F-term scalar potential
\begin{equation}
V = F_i^* F_i
  = (\partial_i W)^* \partial_i W
\end{equation}
is positive definite, so we have the criteria for SUSY breaking:
\begin{equation}
\text{SUSY breaking} \ \Leftrightarrow \ \langle V \rangle > 0
                     \ \Leftrightarrow \ \langle F_i \rangle \ne 0
                     \ \Leftrightarrow \ \langle \partial_i W \rangle \ne 0. \label{eq:3-1-02}
\end{equation}
The full vacuum structure can be obtained from $V$ by solving the equations for stationary points
\begin{equation}
\partial_i V = (\partial_j W)^* \partial_i \partial_j W
             = 0,
\end{equation}
and determining whether a stationary point is a local minimum, maximum or saddle point by checking the eigenvalues of the Hessian matrix
\begin{equation}
\partial^2 V
= \begin{pmatrix}
  \partial_{\bar i} \partial_j V & \partial_{\bar i} \partial_{\bar j} V\\
  \partial_i \partial_j V        & \partial_i \partial_{\bar j} V
  \end{pmatrix}
= \begin{pmatrix}
  (\partial_i \partial_k W)^* \partial_j \partial_k W & (\partial_i \partial_j \partial_k W)^* \partial_k W\\
  (\partial_k W)^* \partial_i \partial_j \partial_k W & (\partial_j \partial_k W)^* \partial_i \partial_k W
  \end{pmatrix}.
\end{equation}
A positive semi-definite Hessian matrix $\partial^2 V \ge 0$ ensures a perturbatively metastable minimum.  It may be a global minimum of $V$, or a local minimum which may decay to another minimum with lower $\langle V \rangle$ through quantum tunneling.  Since $V$ only involves the scalar component $\phi_i$ of chiral superfields, if we just need to determine whether SUSY is broken or not at the vacuum, we can treat the superpotential as $W(\phi_i)$, a function of scalar fields.

A SUSY breaking vacuum can be either a global minimum or a local minimum of $V$.  A solution to the equations $\partial_i W = 0$ gives a SUSY vacuum at the global minimum,  but does not exclude the possibility to have metastable SUSY breaking at some local minimum.  On the other hand, if there is no solution to $\partial_i W = 0$, a minimum of $V$ is always SUSY breaking, although the existence of at least one minimum has to be taken as an assumption.  These possibilities will be demonstrated in the following examples.

\subsection{Examples of Wess-Zumino models}

The Polonyi model~\cite{Polonyi:1977pj}
\begin{equation}
W = f X \label{eq:3-2-01}
\end{equation}
gives a SUSY breaking vacuum at the global minimum
\begin{equation}
X \in \mathbb{C}, \quad
\text{with} \
V = \lvert a \rvert^2,
\end{equation}
with a flat direction along $X$.  As the simplest SUSY breaking model with only one field, it is often used as an effective description of SUSY breaking in the hidden sector.

The deformed Polonyi model~\cite{Intriligator:2007cp}
\begin{equation}
W = f X + \frac{1}{2} h X \phi^2 \label{eq:3-2-02}
\end{equation}
gives two SUSY vacua at the global minima
\begin{equation}
X = 0, \quad
\phi = \pm \sqrt{- \frac{2 f}{h}}, \quad
\text{with} \
V = W
  = 0.
\end{equation}
Either of the two vacua above has no flat direction.  There is also a SUSY breaking stationary point with a flat direction along $X$:
\begin{equation}
X \in \mathbb{C}, \quad
\phi = 0, \quad
\text{with} \
V = \lvert f \rvert^2.
\end{equation}
It is a saddle point in the range $\lvert X \rvert < \lambda$ and becomes a local minimum in the range $\lvert X \rvert > \lambda$.  The transition happens at
\begin{equation}
\lvert X \rvert = \lambda
                = \sqrt{\left \lvert \frac{f}{h} \right \rvert}.
\end{equation}

The original O'Raifeartaigh model~\cite{ORaifeartaigh:1975nky}
\begin{equation}
W = f X_1 + m X_2 \phi + \frac{1}{2} h X_1 \phi^2 \label{eq:3-2-03}
\end{equation}
gives a SUSY breaking vacuum at the global minimum
\begin{equation}
X_1 \in \mathbb{C}, \quad
X_2 = \phi
    = 0, \quad
\text{with} \
V = \lvert f \rvert^2,
\end{equation}
when the coefficients satisfy $\lvert m \rvert^2 > \lvert f h \rvert$.  Note that this vacuum has a flat direction along $X_1$.  When $\lvert m \rvert^2 < \lvert f h \rvert$ is satisfied, the solution above becomes a saddle point, and there appear two new SUSY breaking vacua at the global minima
\begin{equation}
\begin{gathered}
X_1 \in \mathbb{C}, \quad
X_2 = - \frac{h}{m} X_1 \phi, \quad
\phi = \pm i \sqrt{\frac{2 f}{h}
                   \left ( 1 - \left \lvert \frac{m^2}{f h} \right \rvert \right )},\\
\text{with} \
V = \left \lvert \frac{f m^2}{h} \right \rvert
    \left ( 2 - \left \lvert \frac{m^2}{f h} \right \rvert \right ).
\end{gathered}
\end{equation}
Either of the two vacua above has a one complex dimensional flat direction in the $X_1$-$X_2$ space.

The mezzo-O'Raifeartaigh model~\cite{Ellis:1982vi}
\begin{equation}
W = h X_1 \phi (\phi - m_1) + m_2 X_2 (\phi - m_1) \label{eq:3-2-04}
\end{equation}
gives a SUSY vacuum at the global minimum
\begin{equation}
X_1 \in \mathbb{C}, \quad
X_2 = - \frac{h m_1}{m_2} X_1, \quad
\phi = m_1, \quad
\text{with} \
V = W
  = 0.
\end{equation}
It has a one complex dimensional flat direction in the $X_1$-$X_2$ space.  There is no SUSY breaking stationary point when the coefficients satisfy $\lvert h m_1 \rvert^2 < 8 \lvert m_2 \rvert^2$.  When $\lvert h m_1 \rvert^2 > 8 \lvert m_2 \rvert^2$ is satisfied, there appears a metastable SUSY breaking vacuum at the local minimum
\begin{equation}
\begin{gathered}
X_1 \in \mathbb{C}, \quad
X_2 = - \frac{h}{m_2} X_1 (2 \phi - m_1), \quad
\phi = \frac{m_1}{4}
       \left ( 1 - \sqrt{1 - 8 \left \lvert \frac{m_2}{h m_1} \right \rvert^2} \right ),\\
\text{with} \
V = \frac{\lvert h m_1^2 \rvert^2}{32}
    \left ( 1
            + 20 \left \lvert \frac{m_2}{h m_1} \right \rvert^2
            - 2 \left ( 1 - 2 \left \lvert \frac{m_2}{h m_1} \right \rvert^2 \right )
                \sqrt{1 - 8 \left \lvert \frac{m_2}{h m_1} \right \rvert^2}
    \right ),
\end{gathered}
\end{equation}
and a SUSY breaking saddle point at
\begin{equation}
\begin{gathered}
X_1 \in \mathbb{C}, \quad
X_2 = - \frac{h}{m_2} X_1 (2 \phi - m_1), \quad
\phi = \frac{m_1}{4}
       \left ( 1 + \sqrt{1 - 8 \left \lvert \frac{m_2}{h m_1} \right \rvert^2} \right ),\\
\text{with} \
V = \frac{\lvert h m_1^2 \rvert^2}{32}
    \left ( 1
            + 20 \left \lvert \frac{m_2}{h m_1} \right \rvert^2
            + 2 \left ( 1 - 2 \left \lvert \frac{m_2}{h m_1} \right \rvert^2 \right )
                \sqrt{1 - 8 \left \lvert \frac{m_2}{h m_1} \right \rvert^2}
    \right ).
\end{gathered}
\end{equation}
Either the metastable vacuum or the saddle point above has a one complex dimensional flat direction in the $X_1$-$X_2$ space.

The Witten's runaway model~\cite{Witten:1981kv}
\begin{equation}
W = f X + \frac{1}{2} h X^2 \phi \label{eq:3-2-05}
\end{equation}
gives a SUSY runaway direction along
\begin{equation}
\phi = - \frac{f}{h X}, \quad
X \to 0, \quad
\text{with} \
V \to 0,
\end{equation}
and a SUSY breaking saddle point at
\begin{equation}
X = \phi
  = 0, \quad
\text{with} \
V = \lvert f \rvert^2.
\end{equation} 
The scalar potential has no global or local minimum in this model.

\subsection{Including K\"ahler potentials}

The Wess-Zumino models can be generalized to a SUSY non-linear sigma model, which includes a K\"ahler potential in the target space of chiral superfields.  The Lagrangian density \eqref{eq:3-1-01} is generalized to
\begin{equation}
\mathcal{L} = \int \! \ud^4 \! \theta \,
              K(\bar{\Phi}_i, \Phi_i)
              - \left ( \int \! \ud^2 \! \theta \, W(\Phi_i)
                        + \text{c.c.} \right ), \label{eq:3-3-01}
\end{equation}
which leads to the F-term scalar potential
\begin{equation}
V = K^{\bar{i} j} F_i^* F_j
  = K^{\bar{i} j} (\partial_i W)^* \partial_j W, \quad
K^{\bar{i} j} K_{\bar{i} k} = \delta^j_k, \quad
K_{\bar{i} j} = \partial_{\bar{i}} \partial_j K.
\end{equation}
If we just need to determine whether SUSY is broken or not at the vacuum, the K\"ahler potential $K(\bar{\Phi}_i, \Phi_i)$ can be treated as $K(\bar{\phi}_i, \phi_i)$, a function of scalar fields and their conjugates.  A trivial or minimal $K = \phi_i^* \phi_i$ corresponds to the previous Lagrangian density \eqref{eq:3-1-01}, and a non-minimal $K$ may come from integrating out heavy fields for an effective theory.  $K$ is a real and positive function, which means that the K\"ahler metric $K_{\bar{i} j}$ or $K^{\bar{i} j}$ is positive definite.  So the condition for SUSY breaking is the same as the previous condition \eqref{eq:3-1-02}, although the full vacuum structure is different because of the non-minimal $K$.

\subsection{SUGRA extensions}

The previously introduced SUSY transformations do not depend on spacetime coordinates.  The theory is accordingly called to have global SUSY\@.  We can also consider the gauge theory of local SUSY, or SUGRA~\cite{Nilles:1983ge, Wess:1992cp, Bailin:1994qt, VanNieuwenhuizen:1981ae, Kallosh:2000ve, Freedman:2012zz}, which can serve as a low-energy effective description of moduli stabilization in 4-dimensional flux compactification of string theory.  Here we are not going to give a review on SUGRA\@.  Instead, we take the result of the scalar potential involving chiral superfields after tedious calculation~\cite{Cremmer:1978hn}:
\begin{equation}
V = e^{K / M_\text{Pl}^2}
    (K^{\bar i j} (D_i W)^* D_j W - \frac{3}{M_\text{Pl}^2} W^* W), \label{eq:3-4-01}
\end{equation}
where $M_\text{Pl}^2$ is the scale of quantum gravity or the Planck mass, and $D_i W$ is the K\"ahler derivatives of $W$:
\begin{equation}
D_i W = \partial_i W + \frac{1}{M_\text{Pl}^2} W \partial_i K.
\end{equation}
The criteria for SUSY breaking becomes
\begin{equation}
\text{SUSY breaking in SUGRA} \ \Leftrightarrow \ \langle D_i W \rangle \ne 0.
\end{equation}

There are several differences between vacua in global SUSY and SUGRA\@.  Due to the extra term proportional to $W^* W$ in \eqref{eq:3-4-01}.  SUSY vacua can have either $\langle V \rangle = 0$ or $\langle V \rangle < 0$, and SUSY breaking vacua can have either $\langle V \rangle = 0$, $\langle V \rangle > 0$, or $\langle V \rangle < 0$.  Since the magnitudes of $\langle D_i W \rangle$ and $\langle V \rangle$ can be adjusted independently, and ingredients of quantum gravity have been considered in SUGRA, we can view $\langle V \rangle$ seriously as the vacuum energy density, or the cosmological constant.  Furthermore, a solution to the equations $D_i W = 0$ also satisfies $\partial_i V = 0$, but may not lead to a positive semi-definite Hessian matrix $\partial^2 V$.  The change of $V$ along an arbitrary scalar perturbation $\delta \phi_i$ from the stationary point satisfying $D_i W = 0$ is
\begin{equation}
\begin{split}
\delta V &= 2 \partial_{\bar i} \partial_j V \delta \phi_i^* \delta \phi_j
            + \partial_i \partial_j V \delta \phi_i \delta \phi_j
            + \partial_{\bar i} \partial_{\bar j} V \delta \phi_i^* \delta \phi_j^*\\
         &= 2 e^{K / M_\text{Pl}^2} K^{\bar i j}
              \left ( (D_i D_k W \delta \phi_k
                       - \frac{1}{2 M_\text{Pl}^2}
                         W K_{\bar k i} \delta \phi_k^*)^* \right.\\
         &\hspace{7em}
                      \times
                      (D_j D_k W \delta \phi_k
                       - \frac{1}{2 M_\text{Pl}^2}
                         W K_{\bar k j} \delta \phi_k^*)\\
         &\hspace{7em}
              \left.  - \frac{9}{4 M_\text{Pl}^4}
                        (W K_{\bar k i} \delta \phi_k^*)^*
                        (W K_{\bar k j} \delta \phi_k^*) \right )\\
         &\ge - \frac{9}{2 M_\text{Pl}^4}
                e^{K / M_\text{Pl}^2}
                W^* W K_{\bar i j} \delta \phi_i^* \delta \phi_j.
\end{split}
\end{equation}
The mass of the scalar perturbation $\delta \phi_i$ satisfies the bound
\begin{equation}
m^2 = \frac{\delta V}{2 K_{\bar i j} \delta \phi_i^* \delta \phi_j}
    \ge - \frac{9}{4 M_\text{Pl}^4} e^{K / M_\text{Pl}^2} W^* W
    = \frac{3}{4 M_\text{Pl}^2} V,
\end{equation}
which is identical to the Breitenlohner-Freedman bound on masses of scalar fields in 4-dimensional anti-de Sitter (AdS) spacetime with $V$ to be the cosmological constant~\cite{Breitenlohner:1982bm, Breitenlohner:1982jf}.  So a solution to $D_i W = 0$ always gives a metastable SUSY vacuum in this context.  It is still possible to have metastable local minima somewhere else in the field space.  On the other hand, if there is no solution to $D_i W = 0$, a minimum of $V$ is always SUSY breaking, although the existence of at least one minimum has to be taken as an assumption.

As an example, consider a class of models named no-scale SUGRA~\cite{Cremmer:1983bf, Lahanas:1986uc}, in which the superpotential does not depend on one of the chiral superfields $\rho$.  The superpotential and the K\"ahler potential have the form
\begin{align}
W &= W(\phi_i),\\
K &= - 3 M_\text{Pl}^2 \log (i (\rho^* - \rho) / M_\text{Pl}) + K_1(\phi_i^*, \phi_i).
\end{align}
Plugging them into the scalar potential \eqref{eq:3-4-01} gives
\begin{align}
D_\rho W &= 3 W / (\rho^* - \rho),\\
V &= e^{K / M_\text{Pl}^2} K_1^{\bar i j} (D_i W)^* D_j W.
\end{align}
Additional corrections to $V$ are needed to fix $\rho$ at a finite value.  The vacuum has SUSY breaking and $\langle V \rangle = 0$ if we have $\langle D_\rho W \rangle \propto \langle W \rangle \ne 0$ and $\langle D_i W \rangle = 0$, and a small $\langle D_i W \rangle$ can give a small positive $\langle V \rangle$.  Then it is possible to have both SUSY breaking at a physically allowed scale and a small cosmological constant by tuning parameters in the SUGRA context.  Another special type of models have solutions to satisfy
\begin{equation}
\partial_i W = 0 \
\text{and} \
W = 0 \
\Leftrightarrow \
D_i W = 0 \
\text{and} \
V = 0. \label{eq:3-4-02}
\end{equation}
Such solutions also satisfy $\partial_i V = 0$ and $\partial^2 V \ge 0$.  So they give metastable vacuua with SUSY and a zero cosmological constant, or Minkowski SUSY vacua, which contribute to the low-scale SUSY branch of the string landscape~\cite{Dine:2005gz, Dine:2004is, Dine:2005yq}.

\subsection{Summary of vacua in Wess-Zumino models}

In summary, we have the following criteria for SUSY and SUSY breaking minima in Wess-Zumino models and their SUGRA extensions:
\begin{itemize}
\item For Wess-Zumino models in global SUSY:
      \begin{itemize}
      \item SUSY minima $\Leftrightarrow$ solutions to $\partial_i W = 0$;
      \item SUSY breaking minima $\Leftrightarrow$ solutions to $\partial_i V = 0$, $\partial^2 V \ge 0$ and $\partial_i W \ne 0$;
      \item SUSY breaking at a global minimum $\Leftrightarrow$ equations $\partial_i W = 0$ have no solution, and at least one solution to $\partial_i V = 0$ and $\partial^2 V \ge 0$ exists.
      \end{itemize}
\item For SUGRA extensions of Wess-Zumino models:
      \begin{itemize}
      \item SUSY minima $\Leftrightarrow$ solutions to $D_i W = 0$;
      \item SUSY breaking minima $\Leftrightarrow$ solutions to $\partial_i V = 0$, $\partial^2 V \ge 0$ and $D_i W \ne 0$;
      \item SUSY breaking at a global minimum $\Leftrightarrow$ equations $D_i W = 0$ have no solution, and at least one solution to $\partial_i V = 0$ and $\partial^2 V \ge 0$ exists.
      \item SUSY minima with $V = 0$ $\Leftrightarrow$ solutions to $\partial_i W = 0$ and $W = 0$.
      \end{itemize}
\end{itemize}
These criteria hold for models with arbitrary K\"ahler potentials, which can be either minimal or non-minimal.

\subsection{The pseudomodulus}

In previous examples of global SUSY, all SUSY breaking vacua have flat directions, while SUSY vacua may or may not have flat directions.  This is actually a common feature of Wess-Zumino models with minimal K\"ahler potentials~\cite{Zumino:1981pq, Polchinski:1982qk, Einhorn:1983pk}.  The flat direction $\phi_0$, named the pseudomodulus, is not only massless, but also flat up to an arbitrary order at tree level~\cite{Ray:2006wk, Sun:2008nh}.  Meanwhile, since supercharges are fermionic generators of the super-Poincar\'e algebra, the Goldstone's theorem implies a massless fermionic Goldstone particle $\psi_0$, named the Goldstino, at the SUSY breaking vacua.  And in addition, auxiliary fields can be rearranged by field redefinition of $\phi_i$ so that a unique $F_0$ gets a non-zero expectation value for SUSY breaking.  It can be proved that $\phi_0$, $\psi_0$ and $F_0$ are in one chiral superfield, named the SUSY breaking spurion.  Thus the SUSY breaking sector is often approximated by just this one spurion for phenomenology study.  The pseudomodulus $\phi_0$ can acquire mass either from quantum corrections through the Coleman-Weinberg mechanism~\cite{Coleman:1973jx}, from a non-minimal K\"ahler potential, or from SUGRA generalization of the Wess-Zumino model.  For example, if we introduce to the Polonyi model \eqref{eq:3-2-01} a non-minimal K\"ahler potential
\begin{equation}
K = \phi^* \phi - \frac{c}{4 \Lambda^2} (\phi^* \phi)^2, \quad
\text{with} \
c \in \mathbb{R}, \
c > 0,
\end{equation}
where $\Lambda$ is the cut-off scale of the effective theory.  The scalar potential becomes
\begin{equation}
V = \lvert a \rvert^2 (1 - \frac{c}{\Lambda^2} \phi^* \phi)^{- 1}.
\end{equation}
If we instead generalize the Polonyi model to SUGRA while keep the minimal K\"ahler potential, the scalar potential becomes
\begin{equation}
V = \lvert a \rvert^2 e^{\phi^* \phi / M_\text{Pl}^2}
    \left( (\frac{\phi^* \phi}{M_\text{Pl}^2} - \frac{1}{2})^2 + \frac{3}{4} \right).
\end{equation}
In both cases, the SUSY breaking vacuum is stabilized at $\phi = 0$ without a flat direction.  It is possible to have the vacuum stabilized at $\phi \ne 0$ in models with more fields.

\section{Roles of R-symmetries}

\subsection{R-symmetries and discrete R-symmetries}

The existence of a pseudomodulus on any SUSY breaking vacuum in Wess Zumino models has another implication.  Some parameters in the superpotential need to be precisely tuned in order to obtain a flat direction, otherwise the stationary point becomes a maximum or saddle point~\cite{Sun:2008nh}.  So models with SUSY breaking vacua only lives on a submanifold of the whole parameter space, and a small perturbation to parameters can destroy the vacua.  Even if considering non-minimal K\"ahler potentials or SUGRA generalizations, the size of the allowed parameter region is suppressed by a power of the cut-off scale of the effective theory or the Planck mass~\cite{Sun:2011aq}.  This means that SUSY breaking models are non-generic without imposing additional constrains.  Arguments from counting SUSY equations versus variables or the Witten index also suggest the non-genericity of SUSY breaking global vacua~\cite{Witten:1982df}.  The argument from the pseudomodulus applies to metastable SUSY breaking at local vacua as well.  To construct generic SUSY breaking models, a common technique is to constrain the form superpotentials by R-symmetries, and keep the genericity of parameters respecting R-symmetries.  

For $\mathcal{N} = 1$ SUSY, an R-symmery is a $U(1)$ symmetry whose generator does not commute with supercharges.  Fermionic coordinates transform under the R-symmetry according to 
\begin{equation}
\theta_\alpha \to R(\omega) \theta_\alpha
                  = e^{- i \omega} \theta_\alpha, \quad
\bar{\theta}^{\dot{\alpha}} \to R(\omega) \bar{\theta}^{\dot{\alpha}}
                                = e^{i \omega} \theta, \quad
\omega \in [- \pi, \pi).
\end{equation}
Chiral superfields or scalar fields transform under the R-symmetry like
\begin{equation}
\phi_i \to R(\omega) \phi_i
           = e^{- i r_i \omega} \phi, \label{eq:4-1-01}
\end{equation}
where $r_i$ is the R-charge of $\phi_i$, and its conjugate $\phi_i^*$ has R-charge $- r_i$.  So $\theta_\alpha$ and $\bar{\theta}^{\dot{\alpha}}$ respectively have R-charges $1$ and $- 1$.  From the definition of the Berezin integral \eqref{eq:2-3-01} we know that $\ud \! \theta^\alpha$ and $\ud \! \bar{\theta}^{\dot{\alpha}}$ respectively have R-charges $- 1$ and $1$.  If the model has an R-symmetry, the SUSY Lagrangian density must be R-invariant.  The superpotential term in \eqref{eq:3-1-01} or \eqref{eq:3-3-01} means that $W$ transforms like $\theta \theta$ and has R-charge $r_W = 2$, thus each term of $W$ must have R-charges of fields added up to $2$.  The K\"ahler potential $K$ is R-invariant, thus each term of $K$ must have R-charges of fields added up to $0$.

We can also consider discrete $\mathbb{Z}_n$ R-symmetries, which can be viewed as discrete subgroups of continuous $U(1)$ R-symmetries.  Fields transform in a similar way to \eqref{eq:4-1-01}, but the rotation angle take discrete values $\omega = 2 \pi k / n$ with $k = 0, \dots , n-1$.  They are sometimes the remnant symmetry of broken $U(1)$ R-symmetries.  Such models can be considered to have approximate $U(1)$ R-symmetries.  Another type of discrete R-symmetries are identified as geometric symmetries of the internal space, which are common in string compactification.  

\subsection{The Nelson-Seiberg theorem}

The relation between R-symmetries and SUSY breaking is described by the Nelson-Seiberg theorem and its extensions.  The original Nelson-Seiberg theorem is the following~\cite{Nelson:1993nf}:
\begin{theorem} \label{thm:4-01}
(The Nelson-Seiberg theorem) In a Wess-Zumino model with a generic superpotential, an R-symmetry is a necessary condition, and a spontaneously broken R-symmetry is a sufficient condition for SUSY breaking at the vacuum of a global minimum.
\end{theorem}
The proof of the theorem involves counting the number of equations and variables to determine the existence of solutions for generic functions.  Suppose we have $n$ fields $\{ \phi_i \mid i = 1, \dots , N \}$.  Since the number of SUSY equaions $\partial_i W = 0$ and the number of variables $\phi_i$ are both $N$, SUSY solutions exist for a generic $W$.  A symmetry which is not an R-symmetry does not change the situation since it reduces both the number of equations and the number of variables by a same number.  But an R-symmetry can make a difference by constraining the superpotential $W$ to a special form.  Since $W$ has R-charge $2$, at least one field must have a non-zero R-charge.  One can select such a field, supposed to be $\phi_N$ with $r_N \ne 0$, and make the field redefinition
\begin{equation}
X = \phi_N ^ {2 / r_N}, \quad
Y_i = \phi_i / \phi_N ^ {r_i / r_N}, \quad
i = 1, \dots , N - 1. \label{eq:4-2-01}
\end{equation}
$W$ can be expressed with these redefined fields as
\begin{equation}
W = X f(Y_1, \dots , Y_{N - 1}).
\end{equation}
The equations $\partial_i W = 0$ become
\begin{equation}
f = 0, \quad
X \partial_{Y_i} f = 0, \quad
i = 1, \dots , N - 1.
\end{equation} 
If $\langle X \rangle \ne 0$ is assumed, we need to solve $N$ equations with $N - 1$ variables $Y_i$.  There is no such a solution for a generic function $f$.  A vacuum with $\langle X \rangle \ne 0$, if existing, must be SUSY breaking in a model with a generic $W$.  If we can have $\langle X \rangle = 0$, any solution to $f = 0$ seems to give a SUSY vacuum.  But the field redefinition \eqref{eq:4-2-01} is usually singular at $X = 0$ except for some special R-charge arrangement.  So the existence of such a vacuum is unclear.  Notice that $\langle X \rangle \ne 0$ spontaneously breaks the R-symmetry, and at a vacuum with spontaneous R-symmetry breaking we can always find a field $\phi_N$ with $r_N \ne 0$ and $\langle \phi_N \rangle \ne 0$ to make the field redefinition \eqref{eq:4-2-01}.  In summary, SUSY breaking needs R-symmetries, and spontaneous R-symmetry breaking leads to SUSY breaking.  What we have proved is exactly Theorem \ref{thm:4-01}, the original Nelson-Seiberg theorem.

\subsection{The Nelson-Seiberg theorem revised and generalized}

The singular point of the field redefinition \eqref{eq:4-2-01} embarrasses our discussion on any solution at $\langle X \rangle = 0$.  It is the reason why the necessary and sufficient conditions are stated separately in Theorem \ref{thm:4-01}.  Furthermore, to utilize the sufficient condition, one has to find a vacuum which avoids the singularity.  Thus whether SUSY is broken or not can already be seen from the vacuum solution, and the theorem becomes dispensable.  To overcome these obstacles, a revised and generalized theorem has been presented in literature~\cite{Kang:2012fn, Li:2020wdk}:
\begin{theorem} \label{thm:4-02}
(The Nelson-Seiberg theorem revised and generalized) In a Wess-Zumino model with a generic superpotential, SUSY breaking happens at the vacuum of a global minimum if and only if the superpotential has an R-symmetry, and one of the following conditions is satisfied:
      \begin{itemize}
      \item The superpotential is singular at the origin of the field space.
      \item The superpotential is not singular at the origin of the field space, and $N_2 > N_0$ for any consistent R-charge assignment, where $N_2$ and $N_0$ are respectively the number of R-charge $2$ and R-charge $0$ fields.
      \end{itemize}
\end{theorem}
To prove the theorem, we present the R-symmetric $W$ without doing any field redefinition like \eqref{eq:4-2-01}.  Fields are assorted into three types according to their R-charges:
\begin{align}
r_{X_i} &= r_W
         = 2, \quad
i = 1, \dots , N_2,\\
r_{Y_j} &= 0, \quad
j = 1, \dots , N_0,\\
r_{A_k} &\ne r_W \ \text{and} \ r_{A_k} \ne 0, \quad
k = 1, \dots , N_A.
\end{align}
We first consider a polynomial superpotential, and write down the generic form of $W$ with $r_W = 2$:
\begin{align}
W &= X_i f_i(Y_j) + W_A, \label{eq:4-3-01}\\
\begin{split}
W_A &= \underbrace{\xi_{i j k} X_i X_j A_k}_{r_{A_k} = - r_W}
       + \underbrace{\rho_{i j k} X_i A_j A_k}_{r_{A_j} + r_{A_k} = 0}
       + \underbrace{(\mu_{i j} + \nu_{i j k} Y_k) A_i A_j}_{r_{A_i} + r_{A_j} = r_W}
       + \underbrace{\lambda_{i j k} A_i A_j A_k}_{r_{A_i} + r_{A_j} + r_{A_k} = r_W}\\
    &\quad
       + (\text{non-renormalizable terms}). \label{eq:4-3-02}
\end{split}
\end{align}
If $N_2 \le N_0$ is assumed, all first derivatives of $W_A$ can be set to zero by making $\langle X_i \rangle = \langle A_i \rangle = 0$.  Then a solution to $f_i(Y_j) = 0$ gives a SUSY vacua.  Such a solution exists for a generic $W$ because the number of equations $N_2$ is less than or equal to the number of variables $N_0$.  If there are more than one R-charge assignments, one can try to assign different R-charges to fields and satisfy $N_2 \le N_0$.  If $N_2 > N_0$ is satisfied for any consistent R-charge assignment, there is generically no SUSY solution with $\langle X_i \rangle = \langle A_k \rangle = 0$.  In this case, if there is some $X_i$ or $A_i$ with a non-zero expectation value, it spontaneously breaks the R-symmetry and generically leads to SUSY breaking via the original Nelson-Seiberg theorem.  Notice also that absence of an R-symmetry generically leads to unbroken SUSY via the original Nelson-Seiberg theorem.  These exhaust all cases with and without R-symmetries.  Therefore the necessary and sufficient condition for SUSY breaking from a polynomial $W$ is that $W$ has an R-symmetry and $N_2 > N_0$ for any consistent R-charge assignment.

For generic superpotentials of any form, if the R-symmetric superpotential is singular at the origin of the field space, a vacuum must avoid the origin to ensure a reliable effective theory calculation. Then the R-symmetry is broken by non-zero vacuum expectation values of some fields, and SUSY is broken according to the original Nelson-Seiberg theorem.  If the R-symmetric superpotential is not singular at the origin, it has a Taylor expansion in variables $X_i$ and $A_i$:
\begin{align}
W &= X_i f_i(Y_j) + W_A,\\
\begin{split}
W_A &= \underbrace{\xi_{i j k}(Y_l) X_i X_j A_k}_{r_{A_k} = - r_W}
       + \underbrace{\rho_{i j k}(Y_l) X_i A_j A_k}_{r_{A_j} + r_{A_k} = 0}
       + \underbrace{\mu_{i j}(Y_k) A_i A_j}_{r_{A_j} + r_{A_k} = r_W}
       + \underbrace{\lambda_{i j k}(Y_l) A_i A_j A_k}_{r_{A_i} + r_{A_j} + r_{A_k} = r_W}\\
    &\quad
       + (\text{terms contain more than three $X$'s or $A$'s}).
\end{split}
\end{align}
The discussion on a vacuum with $\langle X_i \rangle = \langle A_i \rangle = 0$ in the previous case for polynomial $W$ is still valid.  Any nonzero radius of convergence of the Taylor expansion ensures the existence of such a vacuum.  The discussion on a vacuum with $\langle X_i \rangle \ne 0$ or $\langle A_i \rangle \ne 0$ only involves the original Nelson-Seiberg theorem, which does not rely on the Taylor expansion form of $W$.  Notice again that the absence of an R-symmetry generically leads to unbroken SUSY via the original Nelson-Seiberg theorem.  Summarizing all the discussions, we obtain the necessary and sufficient condition for SUSY breaking, as is stated in Theorem \ref{thm:4-02}.

Several remarks are to be addressed.  Firstly, the possibility to have different consistent R-charge assignments is equivalent to the existence of an extra $U(1)$ symmetry~\cite{Komargodski:2009jf}.  The equivalent $U(1)$ charge $q_i$ is related to different choices of R-charges $r_i$ and $r'_i$ by $r'_i - r_i = \lambda q_i$, where $\lambda$ is an arbitrary non-zero real number.  Secondly, the existence of a minimum must be taken as an assumption.  Models like the Witten's runaway model \eqref{eq:3-2-05}, although satisfying $N_2 > N_0$, have no metastable minimum.  Thirdly, requiring renormalizability may cause non-genericity, since a renormalizable $W$ must be a cubic polynomial function, and the limited number of R-charge $2$ combinations of fields up to cubic may lead to a non-generic form of $W$ which violates the theorem.  And finally, in the non-singular and $N_2 \le N_0$ case, since all field expectation values except the R-invariant ones are set to zero, the R-symmetry is also preserved at the SUSY vacuum.  Therefore there is no contradiction between Theorem \ref{thm:4-01} and Theorem \ref{thm:4-02}.

\subsection{R-symmetric and R-symmetry breaking SUSY vacua}

The R-symmetric vacua in the non-singular and $N_2 \le N_0$ case in the previous proof are especially interesting.  A solution to $\langle X_i \rangle = \langle A_i \rangle = 0$ and $f_i(Y_j) = 0$ gives a vacuum which not only preserves SUSY, but also satisfies $\langle W \rangle = 0$.  This result is also true for models with a discrete $\mathbb{Z}_{n \ge 3}$ R-symmetry, in which $r_W$ is understood as transforming in the same way as $W$ does, and R-charge $0$ is understood as being R-invariant.  So $X$'s transform under the discrete R-symmetry like $W$, $Y$'s are R-invariant, and $A$'s are neither of these two types.  Conditions under the braces in \eqref{eq:4-3-02} require certain transformation properties of field combinations.  Among them, $- r_W$ is understood as transforming in the opposite way as $W$.  A generic $W$ respecting a $\mathbb{Z}_{n \ge 3}$ R-symmetry has the same form as \eqref{eq:4-3-01} and \eqref{eq:4-3-02}.  Even if non-renormalizable terms are considered, each term of $W_A$ must contain at least two $X$'s or $A$'s.  So all first derivatives of $W_A$ can be set to zero by making $\langle X_i \rangle = \langle A_i \rangle = 0$.  The previous argument works the same and gives R-symmetric SUSY vacua with $\langle W \rangle = 0$.  On the other hand, a $\mathbb{Z}_2$ R-symmetry or R-parity keeps $W$ invariant.  There is no distinction between $X$'s and $Y$'s.  Moreover, $\mathbb{Z}_2$ can not be a true R-symmetry since the transformation on supercharges can be absorbed in a Lorentz rotation.  So a $\mathbb{Z}_2$ R-symmetry does not lead to SUSY vacua as in the $\mathbb{Z}_{n \ge 3}$ R-symmetry case.  The sufficient condition for R-symmetric SUSY vacua is summarized as the following~\cite{Sun:2011fq}:
\begin{theorem} \label{thm:4-03}
(A sufficient condition for R-symmetric SUSY vacua) In a Wess-Zumino model with a generic polynomial superpotential, R-symmetric SUSY vacua with $\langle W \rangle = 0$ exist if the superpotential has a $U(1)$ or $\mathbb{Z}_{n \ge 3}$ R-symmetry with $N_2 \le N_0$ for a possible R-charge assignment, where $N_2$ is the number of fields which transform in the same way as the superpotential, and $N_0$ is the number of R-invariant fields.
\end{theorem}

The case $N_2 > N_0$ generically leads to SUSY breaking.  But counterexamples with generic superpotential coefficients are also found~\cite{Sun:2019bnd, Amariti:2020lvx}.  These counterexamples can be viewed as having non-generic R-charges, but are more properly described as possessing special R-charge assignments satisfying certain conditions.  To see such a sufficient condition, we identify $P$'s and $Q$'s from $A$'s in the field classification:
\begin{equation}
\begin{gathered}
r_{P_{(r) i}} = - r_{Q_{(- r) j}}
              = r, \quad
r \ne r_W \ \text{and} \ r \ne 0, \quad
i = 1, \dots , N_{+ r}, \quad
j = 1, \dots , N_{- r},\\
\text{$P$'s and $Q$'s appear in $W_A$ only in terms with at least two $X$'s or $A$'s,} \label{eq:4-4-01}
\end{gathered}
\end{equation}
and keep the rest of $A$'s not satisfying the condition still as $A$'s.  The generic form of $W$ becomes
\begin{align}
W &= X_i f_i(Y_j, P_{(r) j} Q_{(- r) k}) + W_A,\\
\begin{split}
W_A &= \underbrace{\xi_{i j k} X_i X_j A_k}_{r_{A_k} = - r_W}
       + \underbrace{\rho_{i j k} X_i A_j A_k}_{r_{A_j} + r_{A_k} = 0}
       + \underbrace{\sigma_{(r) i j k} P_{(r) i} A_j A_k}_{r_{A_j} + r_{A_k} = r_W - r}
       + \underbrace{\tau_{(r) i j k} Q_{(- r) i} A_j A_k}_{r_{A_j} + r_{A_k} = r_W + r}\\
    &\quad
       + \underbrace{(\mu_{i j} + \nu_{i j k} Y_k) A_i A_j}_{r_{A_i} + r_{A_j} = r_W}
       + \underbrace{\lambda_{i j k} A_i A_j A_k}_{r_{A_i} + r_{A_j} + r_{A_k} = r_W}\\
    &\quad
       + (\text{non-renormalizable terms}).
\end{split}
\end{align}
All first derivatives of $W_A$ can be set to zero by making $\langle X_i \rangle = \langle A_i \rangle = 0$.  Then a solution to $f_i(Y_j, P_{(r) j} Q_{(- r) k}) = 0$ gives a SUSY vacuum.  Such a solution generically exists if $N_2 \le N_0 + N_\pm$ is satisfied, where
\begin{equation}
N_\pm = \sum_r (N_{r} + N_{- r} - 1) \label{eq:4-4-02}
\end{equation}
is the number of independent $P Q$ products in the functions $f_i(Y_j, P_{(r) j} Q_{(- r) k})$.  Since $P$'s and $Q$'s have nonzero R-charges, and a solution to $f_i(Y_j, P_{(r) j} Q_{(- r) k}) = 0$ generically gives them nonzero vacuum expectation values, the SUSY breaking vacuum also breaks the R-symmetry.  In summary, we have obtained the sufficient condition for R-symmetry breaking SUSY vacua~\cite{Sun:2021svm}:
\begin{theorem} \label{thm:4-04}
(A sufficient condition for R-symmetry breaking SUSY vacua) In a Wess-Zumino model with a generic polynomial superpotential, R-symmetry breaking SUSY vacua with $\langle W \rangle = 0$ exist if the superpotential has an R-symmetry with $N_0 < N_2 \le N_0 + N_\pm$ for a possible R-charge assignment, where $N_2$ and $N_0$ are respectively the number of R-charge $2$ and R-charge $0$ fields, and $N_\pm$, given by \eqref{eq:4-4-02}, is the number of independent products of oppositely R-charged fields described by \eqref{eq:4-4-01}.
\end{theorem}
Unlike Theorem \ref{thm:4-03}, this sufficient condition does not have a simple generalization to discrete R-symmetries.  The R-symmetry breaking feature of SUSY vacua implies that they are counterexamples to the original Nelson-Seiberg theorem with non-generic R-charges satisfying Theorem \ref{thm:4-04}.  Most currently known counterexamples to Theorem \ref{thm:4-01} and Theorem \ref{thm:4-02} with generic superpotential coefficients can be described by Theorem \ref{thm:4-04}.  Furthermore, it can be proved that models satisfying Theorem \ref{thm:4-03} and Theorem \ref{thm:4-04} always give SUSY vacua.  There is no special R-charge assignment which can give SUSY breaking counterexample with generic coefficients~\cite{Li:2021ydn}.  But new counterexamples with R-symmetry breaking SUSY vacua not covered by the sufficient condition are also found~\cite{Brister:2022rrz}.

As we have shown before in \eqref{eq:3-4-02}, solutions to $\partial_i W = 0$ and $W = 0$ are also solutions to $D_i W = 0$ and $V = 0$ in SUGRA\@.  Therefore Theorem \ref{thm:4-03} and Theorem \ref{thm:4-04} also give SUSY vacua with $\langle V \rangle = 0$ in SUGRA\@.  The feature $\langle W \rangle = 0$ is actually generic for any SUSY vacuum from a superpotential with a $U(1)$ R-symmetry~\cite{Kappl:2008ie}.  The superpotential transforms under an R-symmetry:
\begin{equation}
W \to R(\omega) W
      = e^{- i r_W \omega} W
      = e^{- 2 i \omega} W. \label{eq:4-4-03}
\end{equation}
An infinitesimal transformation of $W$ can be expressed as infinitesimal transformations of fields:
\begin{equation}
\frac{\ud \! W}{\ud \! \omega} = \partial_i W \frac{\ud \! \phi_i}{\ud \! \omega}. \label{eq:4-4-04}
\end{equation}
Since a SUSY vacuum satisfies $\partial_i W = 0$, the consistency between \eqref{eq:4-4-03} and \eqref{eq:4-4-04} means $W = 0$ at the vacuum.  Moreover, it can be proved that an R-symmetric $W$ vanishes term-by term at a SUSY vacuum if its coefficients are generic~\cite{Brister:2021xca}.  Suppose the superpotential is constructed from a set of terms $\{ p_\alpha(\phi_i) \}$ allowed by a certain set of conditions including the R-symmetry:
\begin{equation}
W(\phi_i, c_\alpha) = c_\alpha p_\alpha(\phi_i),
\end{equation}
where $\{ c_\alpha \}$ is a set of coefficients taking generic complex values.  Genericity means that a small perturbation to $c$'s does not affect the existence of a SUSY vacuum with $\langle W \rangle = 0$, although the vacuum expectation values of $\phi$'s are perturbed according to the perturbation of $c$'s.  Thus the notion of genericity can be analytically expressed as
\begin{equation}
\frac{\ud \! W}{\ud \! c_\alpha} = \frac{\partial W}{\partial c_\alpha}
                                   + \partial_i W \frac{\ud \! \phi_i}{\ud \! c_\alpha}
                                 = 0
\end{equation}
at the vacuum.  Since a SUSY vacuum satisfies $\partial_i W = 0$, we have
\begin{equation}
\frac{\partial W}{\partial c_\alpha} = p_\alpha(\phi_i)
                                     = 0,
\end{equation}
which means that every term of $W$ vanishes individually at the SUSY vacuum.

\subsection{A bound on the superpotential}

The feature $\langle W \rangle = 0$ of a SUSY vacuum from an R-symmetric superpotential does not depend on whether the vacuum preserves or breaks the R-symmetry, and does not require any assumption of genericity.  It is the SUSY limit of a more general bound on the superpotential~\cite{Dine:2009sw}:
\begin{theorem} \label{thm:4-05}
(A bound on the superpotential) In $\mathcal{N} = 1$ SUSY theories with SUSY breaking and R-symmetry breaking, the superpotential is bounded by the inequality $\lvert \langle W \rangle \rvert \le f_a f / 2$, where $f_a$ and $f$ are respectively the R-axion decay constant and the Goldstino decay constant.  The inequality is strict for SUSY breaking models interacting with other sectors.
\end{theorem}
When the R-symmetry is broken by non-zero field expectation values, the Goldstone's theorem implies a massless scalar field $a$, names the R-axion.  Scalar fields $\phi_i$ can be parameterized by the R-axion as
\begin{equation}
\phi_i = \lvert \langle \phi \rangle \rvert e^{- i r_i a}.
\end{equation}
Another massless field considered in the low energy effective theory is the Goldstino $\psi_0$ from SUSY breaking.  Using the formulation of non-linearly realized SUSY as well as non-linearly realized R-symmetry, the R-axion kinetic term as well as the Akulov-Volkov terms for the Goldstino are~\cite{Volkov:1972jx, Volkov:1973ix, Deser:1977uq}
\begin{equation}
\begin{split}
\mathcal{L} =& - f^2
               + f_a^2 \partial_\mu a \partial^\mu a
               + i \psi_0^\dagger \bar{\sigma}^\mu \partial_\mu \psi_0
               + \frac{1}{4 f^2} \partial_\mu (\psi_0^\dagger \psi_0^\dagger) \partial^\mu (\psi_0 \psi_0)\\
             & + \text{(interacting terms)},
\end{split}
\end{equation}
where the R-axion decay constant $f_a$ and the Goldstino decay constant $f$ are defined as
\begin{align}
f_a &= \lVert r_i \langle \phi_i \rangle \rVert
     = \sqrt{r_i^2 \langle \phi_i^* \phi_i \rangle},\\
f &= \lVert \langle \partial_i W \rangle \rVert
   = \sqrt{\langle (\partial_i W)^* \partial_i W \rangle}.
\end{align}
In an R-symmetric Wess-Zumino model, an infinitesimal transformation of the R-symmetry gives
\begin{equation}
\frac{\ud}{\ud \! \omega} (R(\omega) W)
= - 2 i W
= \partial_i W \frac{\ud}{\ud \! \omega} (R(\omega) \phi_i)
= - i r_i \phi_i \partial_i W.
\end{equation}
Thus the following identity is satisfied:
\begin{equation}
W = \frac{1}{2} r_i \phi_i \partial_i W.
\end{equation}
Using the Cauchy-Schwarz inequality, we obtain the bound
\begin{equation}
\lvert \langle W \rangle \rvert
= \frac{1}{2} \lvert \langle r_i \phi_i \partial_i W \rangle \rvert
\le \frac{1}{2}
    \sqrt{r_i^2 \langle \phi_i^* \phi_i \rangle
          \langle (\partial_i W)^* \partial_i W \rangle}
= \frac{1}{2} f_a f.
\end{equation}
The equality holds if and only if we have $\langle r_i \phi_i \rangle = \lambda \langle \partial_i W \rangle$ for some $\lambda \in \mathbb{C}$, or $\langle \partial_i W \rangle = 0$.  A more general proof~\cite{Dine:2009sw} includes non-minimal K\"ahler potentials, D-terms, higher derivative corrections to the Lagrangian, etc..  In a general low energy effective field theory restricted by non-linearly realized SUSY and R-symmetry, once radiative corrections are included in an interacting model, the strict inequality always holds.  Note that the SUSY limit $f = 0$ always gives $\langle W \rangle = 0$ even if the R-symmetry is broken and $f_a \ne 0$.  However Theorem \ref{thm:4-05} does not tell the existence of a SUSY vacuum in a model, and only applies to continuous R-symmetries.  In comparison, Theorem \ref{thm:4-03} and Theorem \ref{thm:4-04} provide sufficient conditions for SUSY vacua, and Theorem \ref{thm:4-03} applies also to discrete R-symmetries which are common in string compactification.

\subsection{Summary of R-symmetry related theorems}

We summarize the previous theorems connecting R-symmetries and SUSY breaking with the following short notations:
\begin{itemize}
\item SUSY breaking at global minima $\Rightarrow$ $U(1)^\text{R}$, $\Leftarrow$ spontaneously broken $U(1)^\text{R}$ (requiring a generic $W$);
\item SUSY breaking at global minima $\Leftrightarrow$ $U(1)^\text{R}$ and (1) $W$ is singular at the origin or (2) $W$ is not singular at the origin and $N_2 > N_0$ for any R-charge assignment (requiring a generic $W$);
\item $U(1)^\text{R}$ or $\mathbb{Z}_{n \ge 3}^\text{R}$ and $N_2 \le N_0$ for a possible R-charge assignment $\Rightarrow$ R-symmetric SUSY vacua with $\langle W \rangle = 0$ $\Leftrightarrow$ R-symmetric SUSY vacua in SUGRA with $\langle V \rangle = 0$ (requiring a generic polynomial $W$);
\item $U(1)^\text{R}$ and $N_0 < N_2 \le N_0 + N_\pm$ for a possible R-charge assignment $\Rightarrow$ R-symmetry breaking SUSY vacua with $\langle W \rangle = 0$ $\Leftrightarrow$ R-symmetry breaking SUSY vacua in SUGRA with $\langle V \rangle = 0$ (requiring a generic polynomial $W$);
\item SUSY breaking and $U(1)^\text{R}$ breaking $\Rightarrow$ $\lvert \langle W \rangle \rvert \le f_a f / 2$ (no need for genericity);
\end{itemize}
These theorems hold for models with arbitrary K\"ahler potentials which can be either minimal or non-minimal, and no constraint of R-symmetries on the K\"ahler potential is needed.  The following examples will show how these theorems work.

\subsection{Examples of R-symmetric Wess-Zumino models}

The Polonyi model~\cite{Polonyi:1977pj} has only one field with the R-charge
\begin{equation}
r_1 = 2,
\end{equation}
and thus $N_2 = 1 > N_0 = 0$.  The superpotential
\begin{equation}
W = a \phi_1
\end{equation}
gives a SUSY breaking vacuum at
\begin{equation}
\phi_1 \in \mathbb{C}, \quad
\text{with} \
V = \lvert a \rvert^2.
\end{equation}
The R-symmetry is broken everywhere in the field space except at the origin.  This model is the same as the previously presented Polonyi model \eqref{eq:3-2-01}.

Adding another R-invariant field, the Polonyi model is modified to have the R-charge assignment
\begin{equation}
\{ r_1, r_2 \} = \{ 2, 0 \},
\end{equation}
and thus $N_2 = 1 = N_0 = 1$.  The superpotential
\begin{equation}
W = a \phi_1 + b \phi_1 \phi_2 + c \phi_1 \phi_2^2 \label{eq:4-7-01}
\end{equation}
gives two R-symmetric SUSY vacua at
\begin{equation}
\phi_1 = 0, \quad
\phi_2 = \frac{1}{2 c} (- b \pm \sqrt{b^2 - 4 a c}) \quad
\text{with} \
V = W
  = 0.
\end{equation}
There is also a SUSY breaking stationary point at
\begin{equation}
\phi_1 \in \mathbb{C}, \quad
\phi_2 = - \frac{b}{2 c}, \quad
\text{with} \
V = \left \lvert a - \frac{b^2}{4 c} \right \rvert^2.
\end{equation}
It is a saddle point in the range $\lvert \phi_1 \rvert < \lambda$ and becomes a local minimum in the range $\lvert \phi_1 \rvert > \lambda$.  The transition happens at
\begin{equation}
\lvert \phi_1 \rvert = \lambda
                     = \frac{\sqrt{\lvert b^2 - 4 a c \rvert}}{2 \sqrt{2} \lvert c \rvert}.
\end{equation}
The R-symmetry is broken everywhere in the flat direction along $\phi_1$ except at the origin.  Note that the previously presented deformed Polonyi model \eqref{eq:3-2-02} is a special case of this models with $b = 0$.

The general O'Raifeartaigh model~\cite{Intriligator:2007cp} has three fields with the R-charge assignment
\begin{equation}
\{ r_1, r_2, r_3 \} = \{ 2, 2, 0 \},
\end{equation}
and thus $N_2 = 2 > N_0 = 1$.  The superpotential
\begin{equation}
W = a \phi_1 + b \phi_2 + c \phi_1 \phi_3 + d \phi_2 \phi_3 + e \phi_1 \phi_3^2 + f \phi_2 \phi_3^2
\end{equation}
gives several SUSY breaking vacua and saddle points by solving the equations
\begin{gather}
\phi_1 (c + 2 e \phi_3) + \phi_2 (d + 2 f \phi_3) = 0,\\
(a + c \phi_3 + e \phi_3^2)^* (c + 2 e \phi_3) + (b + d \phi_3 + f \phi_3^2)^* (d + 2 f \phi_3) = 0.
\end{gather}
At any of the stationary points, the R-symmetry is broken everywhere in the flat direction in the $\phi_1$-$\phi_2$ space except at the origin.  There is no SUSY stationary point in this model with generic coefficients.  Note that the original O'Raifeartaigh model \eqref{eq:3-2-03} is a special case of this models with $b = c = f = 0$.  The previously presented mezzo-O'Raifeartaigh model \eqref{eq:3-2-04} is also a special case which gives a SUSY vacuum with non-generic coefficients $\{ a, b, c, d, e, f \} = \{ 0, - m_1 m_2, - h m_1, m_2, h, 0 \}$.

The metastable SUSY breaking model~\cite{Shih:2007av} has four fields with the R-charge assignment
\begin{equation}
\{ r_1, r_2, r_3, r_4 \} = \{ 2, 1, - 1, 3 \},
\end{equation}
and thus $N_2 = 1 > N_0 = 0$.  The superpotential
\begin{equation}
W = a \phi_1 + b \phi_2^2 + c \phi_1 \phi_2 \phi_3 + d \phi_3 \phi_4 \label{eq:4-7-02}
\end{equation}
gives a SUSY breaking stationary point at
\begin{equation}
\phi_1 \in \mathbb{C}, \quad
\phi_2 = \phi_3
       = \phi_4
       = 0, \quad
\text{with} \
V = \lvert a \rvert^2.
\end{equation}
It is a saddle point for all values of $\phi_1 \in \mathbb{C}$ when the coefficients satisfy $\lvert a c \rvert > 2 \lvert b d \rvert$.  When $\lvert a c \rvert < 2 \lvert b d \rvert$ is satisfied, it is a local minimum in the range $\lvert \phi_1 \rvert < \lambda$ and becomes a saddle point in the range $\lvert \phi_1 \rvert > \lambda$.  The transition happens at
\begin{equation}
\lvert \phi_1 \rvert = \lambda
                     = \frac{4 \lvert b d \rvert^2 - \lvert a c \rvert^2}{4 \lvert a b c^2 \rvert}.
\end{equation}
The R-symmetry is broken everywhere in the flat direction along $\phi_1$ except at the origin.  There is also a SUSY runaway direction along
\begin{equation}
\phi_1 = \frac{2 a b}{c^2 \phi_3^2}, \quad
\phi_2 = - \frac{a}{c \phi_3}, \quad
\phi_4 = \frac{2 a^2 b}{c^2 d \phi_3^3}, \quad
\phi_3 \to 0, \quad
\text{with} \
V \to 0.
\end{equation}

The Witten's runaway model~\cite{Witten:1981kv} has two fields with the R-charge assignment
\begin{equation}
\{ r_1, r_2 \} = \{ 2, - 2 \},
\end{equation}
and thus $N_2 = 1 > N_0 = 0$.  The superpotential
\begin{equation}
W = a \phi_1 + b \phi_1^2 \phi_2 \label{eq:4-7-03}
\end{equation}
gives a SUSY runaway direction along
\begin{equation}
\phi_2 = - \frac{a}{2 b \phi_1}, \quad
\phi_1 \to 0, \quad
\text{with} \
V \to 0,
\end{equation}
and an R-symmetric SUSY breaking saddle point at
\begin{equation}
\phi_1 = \phi_2
       = 0, \quad
\text{with} \
V = \lvert a \rvert^2,
\end{equation}
But no global or local minimum exists in this model.  This model is the same as the previously presented Witten's runaway model \eqref{eq:3-2-05} but with a different coefficient assignment.

The simplest counterexample model~\cite{Sun:2019bnd} satisfying Theorem \ref{thm:4-04} has four fields with the R-charge assignment
\begin{equation}
\{ r_1, r_2, r_3, r_4 \} = \{ 2, -2, 6, -6 \},
\end{equation}
and thus $N_2 = 1 > N_0 = 0$.  The superpotential
\begin{equation}
W = a \phi_1 + b \phi_1^2 \phi_2 + c \phi_2^2 \phi_3 + d \phi_1 \phi_3 \phi_4 \label{eq:4-7-04}
\end{equation}
gives an R-symmetry breaking SUSY vacuum at
\begin{equation}
\phi_1 = \phi_2
       = 0, \quad
\phi_4 = - \frac{a}{d \phi_3}, \quad
\phi_3 \in \mathbb{C}, \quad
\text{with} \
V = W
  = 0,
\end{equation}
a SUSY runaway direction along
\begin{equation}
\begin{gathered}
\phi_1 = - \frac{c}{d} u v, \quad
\phi_2 = \frac{v}{u}, \quad
\phi_3 = (\frac{2 b c}{d^2} v^2 - \frac{a}{d}) \frac{u^3}{v}, \quad
\phi_4 = \frac{v}{u^3},\\
v = \sqrt \frac{a d (10 + 2 \lvert u \rvert^4)}{b c (25 + 4 \lvert u \rvert^4)}, \quad
u \to 0, \quad
\text{with} \
V \to 0,
\end{gathered}
\end{equation}
and an R-symmetric SUSY breaking saddle point at
\begin{equation}
\phi_1 = \phi_2
       = \phi_3
       = \phi_4
       = 0, \quad
\text{with} \
V = \lvert a \rvert^2.
\end{equation}
There are also R-symmetry breaking SUSY breaking saddle points found in numerical calculation with typical coefficient values.  Note that $\phi_3$ and $\phi_4$ satisfy the condition \eqref{eq:4-4-01} to be identified as $P_{(6)}$ and $Q_{(- 6)}$ fields.  So we have $N_2 = N_0 + N_\pm = 1$ and this model satisfied the sufficient condition in Theorem \ref{thm:4-04}.

As an example of R-charge ambiguity~\cite{Komargodski:2009jf}, consider three fields with the R-charge assignment
\begin{equation}
\{ r_1, r_2, r_3 \} = \{ 2, r, -r \}, \quad
r \in \mathbb{R},
\end{equation}
or equivallently, the above R-charges with a fixed $r$ and an extra $U(1)$ symmetry with charges
\begin{equation}
\{ q_1, q_2, q_3 \} = \{ 0, q, -q \}, \quad
q \in \mathbb{R}.
\end{equation}
Choosing $r = 0$, we have $N_2 = 1 < N_0 = 2$.  The superpotential
\begin{equation}
W = a \phi_1 + b \phi_1 \phi_2 \phi_3 \label{eq:4-7-05}
\end{equation}
gives an R-symmetric SUSY vacuum at
\begin{equation}
\phi_1 = 0, \quad
\phi_2 \phi_3 = - \frac{a}{b}, \quad
\text{with} \
V = W
  = 0.
\end{equation}
If $r \ne 0$ is chosen, the R-symmetry is broken everywhere in the flat direction in the $\phi_2$-$\phi_3$ space.

All R-symmetric models presented here include all R-charge $2$ terms up to cubic, with generic coefficients.  And all R-charges except the ones in the R-charge ambiguity model are uniquely fixed by requiring each term of $W$ to have R-charge $2$.  The theorems in this section give correct predictions of the vacua at the global minima in all the above models except the Witten's runaway model which has no metastable minimum.

\subsection{Runaway directions}

Theorem \ref{thm:4-01} and Theorem \ref{thm:4-02} require the existence of a minimum as an assumption, and Theorem \ref{thm:4-03} and Theorem \ref{thm:4-04} describe SUSY vacua as the global minima of the model.  The full vacuum structure may have metastable local minima which can decay to other minima with lower $\langle V \rangle$ through quantum tunneling.  Another possibility is to have runaway directions like the ones in the metastable SUSY breaking model \eqref{eq:4-7-02} and the Witten's runaway model \eqref{eq:4-7-03}.  Along a runaway direction, some field values goes to infinity, and the scalar potential keeps decreasing toward a asymptotic limit.  The runaway direction can be SUSY where the asymptotic limit of $V$ is zero, or SUSY breaking where the asymptotic limit of $V$ is greater than zero.

It has been found that SUSY runaway directions are common in SUSY breaking models constructed with R-symmetries~\cite{Ferretti:2007ec, Ferretti:2007rq}.  Many of them are associated with complexified R-symmetries, under which fields transform like
\begin{equation}
\phi_i \to R(i \alpha) \phi_i
           = e^{r_i \alpha} \phi, \quad
\alpha \in \mathbb{R},
\end{equation}
and F-terms transform like
\begin{equation}
\partial_i W \to R(i \alpha) \partial_i W
                 = e^{(r_W - r_i) \alpha} W
                 = e^{(2 - r_i) \alpha} W.
\end{equation}
SUSY equations can be classified according to their F-term R-charges:
\begin{align}
\partial_i W &= 0, \quad
r_i > 2 \ \text{and} \ r_{\partial_i W} < 0, \label{eq:4-8-01}\\
\partial_i W &= 0, \quad
r_i = 2 \ \text{and} \ r_{\partial_i W} = 0, \label{eq:4-8-02}\\
\partial_i W &= 0, \quad
r_i < 2 \ \text{and} \ r_{\partial_i W} > 0. \label{eq:4-8-03}
\end{align}
For SUSY breaking models, these equations can not be solved simultaneously.  If one can just solve \eqref{eq:4-8-01} and \eqref{eq:4-8-02}, the complexified R-transformation with $\alpha \to - \infty$ asymptotically satisfies \eqref{eq:4-8-03} and gives a SUSY runaway direction.  The Witten's runaway model \eqref{eq:4-7-03} is an example of this case.  If one can just solve \eqref{eq:4-8-02} and \eqref{eq:4-8-03}, the complexified R-transformation with $\alpha \to + \infty$ asymptotically satisfies \eqref{eq:4-8-01} and gives a SUSY runaway direction.  The metastable SUSY breaking model \eqref{eq:4-7-02} is an example of this case.  If \eqref{eq:4-8-02} can not be solved, one needs to minimize part of the scalar potential from R-charge $0$ F-terms:
\begin{equation}
V_0 = \sum_{r_i = 2} (\partial_i W)^* \partial_i W.
\end{equation}
The complexified R-transformation leaves these F-terms invariant.  One may get SUSY breaking runaway directions if one of \eqref{eq:4-8-01} or \eqref{eq:4-8-03} can be solved, and the other can be asymptotically satisfied by a complexified R-transformation similar to the SUSY case.  Examples of such SUSY breaking runaway directions appear in literature~\cite{Ferretti:2007ec, Ferretti:2007rq}.

The technique of complexified transformation can also be applied to $U(1)$ symmetries which are not R-symmetries.  The complexified $U(1)$ transformation gives
\begin{equation}
\phi_i \to e^{q_i \alpha} \phi, \quad
\alpha \in \mathbb{R}
\end{equation}
and
\begin{equation}
\partial_i W \to e^{- q_i \alpha} W,
\end{equation}
where $q_i$ is the $U(1)$ charge of $\phi_i$.  For an R-symmetric Wess-Zumino model with an extra $U(1)$ symmetry, SUSY equations can be classified according to their F-term $U(1)$ charges:
\begin{align}
\partial_i W &= 0, \quad
q_i > 0 \ \text{and} \ q_{\partial_i W} < 0, \label{eq:4-8-04}\\
\partial_i W &= 0, \quad
q_i = 0 \ \text{and} \ q_{\partial_i W} = 0, \label{eq:4-8-05}\\
\partial_i W &= 0, \quad
q_i < 0 \ \text{and} \ q_{\partial_i W} > 0. \label{eq:4-8-06}
\end{align}
For SUSY breaking models, these equations can not be solved simultaneously.  If one can just solve \eqref{eq:4-8-04} and \eqref{eq:4-8-05}, the complexified $U(1)$ transformation with $\alpha \to - \infty$ asymptotically satisfies \eqref{eq:4-8-06} and gives a SUSY runaway direction.  If one can just solve \eqref{eq:4-8-05} and \eqref{eq:4-8-06}, the complexified $U(1)$ transformation with $\alpha \to + \infty$ asymptotically satisfies \eqref{eq:4-8-04} and gives a SUSY runaway direction.  If \eqref{eq:4-8-05} can not be solved, one needs to minimize part of the scalar potential from F-terms with $U(1)$ charge $0$:
\begin{equation}
V_0 = \sum_{q_i = 0} (\partial_i W)^* \partial_i W.
\end{equation}
The complexified $U(1)$ transformation leaves these F-terms invariant.  One may get SUSY breaking runaway directions if one of \eqref{eq:4-8-04} or \eqref{eq:4-8-06} can be solved, and the other can be asymptotically satisfied by a complexified $U(1)$ transformation similar to the SUSY case.  Examples of runaway directions associated with complexified $U(1)$ symmetries can be found in literature~\cite{Azeyanagi:2012pc, Sun:2018hnk}.

There might be other types of runaway directions which are not associated with R-symmetries or $U(1)$ symmetries.  Since $V$ has $0$ as its lower bound, models which do not process a minimum must have runaway directions.  If corrections at large field values are introduced to the scalar potential, the runaway direction can be lifted up and a SUSY breaking vacuum can be constructed~\cite{Azeyanagi:2012pc}.  By this means, the nonexistence of a solution to the SUSY equations can be taken seriously as a criteria for SUSY breaking, and the assumption of the existence of a minimum in Theorem \ref{thm:4-01} and Theorem \ref{thm:4-02} can be dropped.

\subsection{Applications of R-symmetric Wess-Zumino models}

Applications of R-symmetries using the previous theorems fall into two categories.  On the phenomenology side, R-symmetries are commonly used to build SUSY breaking models.  One can start from an R-symmetric polynomial $W$ with $N_2 > N_0$, or a non-perturbative $W$ with the vacuum known to break the R-symmetry.  The vacuum at the global minimum, if existing, breaks SUSY according to Theorem \ref{thm:4-01} or Theorem \ref{thm:4-02}.  Coupling the SUSY breaking sector to other sectors may introduce R-symmetry breaking corrections to $W$, and SUSY vacua may emerge.  But one can imagine that SUSY breaking vacua are still metastable local minima and long-lived against the Coleman-de~Luccia decay~\cite{Coleman:1977py, Callan:1977pt, Coleman:1980aw} if R-symmetry breaking corrections are small.  Thus metastable SUSY breaking Wess-Zumino models are build with approximate R-symmetries~\cite{Intriligator:2007py, Abe:2007ax}.  Such models can be viewed as low-energy effective descriptions of dynamical SUSY breaking models, such as the ones from Seiberg duality~\cite{Intriligator:2006dd}.  As a bonus, Theorem \ref{thm:4-02} allows arbitrary number of $A$'s with R-charges other than $2$ and $0$, which do not affect the SUSY breaking vacua when $N_2 > N_0$ is satisfied.  These $A$'s are required for spontaneous R-symmetry breaking which generates the Majorana masses of gauginos at one-loop level~\cite{Shih:2007av, Carpenter:2008wi, Sun:2008va, Curtin:2012yu, Liu:2014ida}.

On the string phenomenology side, R-symmetric SUSY vacua from Theorem \ref{thm:4-03} play important roles because of their property $\langle V \rangle = 0$ in SUGRA extensions of Wess-Zumino models.  Continuous global symmetries are not allowed in full quantum gravity theories such as string theory~\cite{Banks:1988yz, Kallosh:1995hi, Banks:2010zn}, but discrete symmetries are common.  In Type IIB flux compactification~\cite{Grana:2005jc, Douglas:2006es, Blumenhagen:2006ci, Denef:2007pq, Samtleben:2008pe, Lust:2009kp}, string theory in 10-dimensions is compactified on a 6-dimensional Calabi-Yau manifold with quantized fluxes of gauge fields or higher gauge fields.  The low-energy effective theory is described by 4-dimentional $\mathcal{N} = 1$ SUGRA, and non-zero fluxes induce a Gukov-Vafa-Witten superpotential for moduli stabilization~\cite{Gukov:1999ya}.  A vast number of different choices in the theory construction, such as the Calabi-Yau manifold or the fluxes, result in a vast number of metastable vacua with different properties, making up the landscape of vacua~\cite{Giddings:2001yu, Susskind:2003kw, Ashok:2003gk, Denef:2004ze, Denef:2004cf, Hebecker:2020aqr}.  A special subset of the landscape, named the low-energy SUSY branch~\cite{Dine:2004is, Dine:2005yq}, has Calabi-Yau manifolds stabilized at points in the moduli space with enhanced discrete symmetries~\cite{DeWolfe:2004ns, DeWolfe:2005gy}.  One can identify one of these geometrical symmetries as a discrete R-symmetry, turn on only invariant fluxes to get an R-symmetric superpotential.  The condition $N_2 \le N_0$ can be arranged in many models with symmetric Calabi-Yau manifolds, such as the $T^6 / \mathbb{Z}_2$ orientifold~\cite{Kachru:2002he}, the quintic surface in $\mathbb{CP}^4$~\cite{Green:2012pqa}, the hypersurfaces in weighted projective spaces~\cite{Klemm:1992bx, Kreuzer:1992np}, or other classes of Calabi-Yau manifolds, producing a landscape of Minkowski SUSY vacua~\cite{Sun:2011fq, Dine:2005gz}.  Such compactifications are studied from the perspective of arithmetic geometry and geometric modularity~\cite{Kanno:2017nub, Kachru:2020sio, Schimmrigk:2020dfl, Kanno:2020kxr}.  Distributions of the resulting Minkowski SUSY vacua are investigated with persistent homology~\cite{Cole:2018emh, Cole:2019enn}.  And similar constructions also arise from non-geometric compactifications~\cite{Micu:2007rd, Palti:2007pm}.  Non-perturbative corrections, such as the racetrack superpotentials~\cite{Krasnikov:1987jj, Casas:1990qi, Taylor:1990wr, deCarlos:1992kox, Dine:1999dx}, are then included to generate SUSY breaking and the superpotential at hierarchically small scales~\cite{Giryavets:2003vd, Denef:2004dm, Cicoli:2013swa, Demirtas:2019sip, Broeckel:2020fdz, Demirtas:2020ffz, Alvarez-Garcia:2020pxd, Honma:2021klo, Demirtas:2021nlu, Demirtas:2021ote, Broeckel:2021uty, Bastian:2021hpc, Grimm:2021ckh, Carta:2021kpk}.  The constraint from the de Sitter swampland conjecture~\cite{Obied:2018sgi, Garg:2018reu, Ooguri:2018wrx, Palti:2019pca} may be evaded~\cite{Conlon:2018eyr, Blanco-Pillado:2018xyn}, and MSSM or its extensions may hopefully be built on these vacua.

With polynomial superpotentials, the criteria for SUSY breaking in Theorem \ref{thm:4-02} or SUSY vacua in Theorem \ref{thm:4-03} requires just counting and comparing $N_2$ and $N_0$.  Assuming genericity, applying these theorems does not require the explicit form of $W$ or vacuum solutions.  It enables us to quickly survey a big set of different models and select the desired one to continue the explicit model building.  Statistical properties of vacua in the landscape, such as the distribution of the cosmological constant and the SUSY breaking scale, can also be studied by this field-counting method.

\section{A dataset of R-symmetric Wess-Zumino models}

\subsection{Motivation from non-generic counterexamples}

The Nelson-Seiberg theorem and its extensions require genericity of the superpotential.  When applying the field-counting method to the landscape of different models, non-generic models may affect the accuracy of the result.  There are two types of non-genericity in our scope.  The usual concept of gennericity is about parameters or coefficients of a model.  A generic model should include all terms respecting symmetries, and the coefficient of each term should take a generic value.  This type of genericity is related to naturalness~\cite{tHooft:1979rat, Barbieri:1987fn, Anderson:1994dz, Giudice:2008bi, Feng:2013pwa, Dine:2015xga, Dijkstra:2019zfy, Borrelli:2019uab, Craig:2022uua}, meaning that there is no fine-tuning in parameters.  It is also related to hierarchy problems~\cite{Wilson:1970ag, Georgi:1974yf, Gildener:1976ai, Weinberg:1978ym}, where a fine-tuning of cancellation between two large parameters gives a small parameter.  For a certain model, genericity in parameters can be easily checked by adding random perturbations on parameters.  If the key properties of the model, such as the existence of SUSY or SUSY breaking vacua, is unchanged after the perturbation, the model is considered to be generic.  Examples with this type of non-genericity include the mezzo-O’Raifeartaigh model \eqref{eq:3-2-04}, where the SUSY vacuum from a non-generic choice of coefficients can be destroyed by adding a small perturbation to coefficients.  This type of non-generic models can only compose a null set in the parameter space, and can be neglected in the survey or statistical study of the landscape.

Another type of non-genericity comes from the combined requirements of R-symmetries and renormalization~\cite{Sun:2019bnd, Amariti:2020lvx, Sun:2021svm, Li:2021ydn, Brister:2022rrz}.  A renormalizable $W$ has terms up to cubic.  With a certain R-charge assignment, there are only finite number of R-charge $2$ field combinations up to cubic.  Some R-charge assignment may restrict $W$ to a specific form, and the actual vacua from explicit solving equations for minima of $V$ may contradict what Theorem \ref{thm:4-01} and Theorem \ref{thm:4-02} predict.  We view these models as counterexamples with non-generic R-charges, so they do not contradict the previous theorems.  Examples with this type of non-genericity include the simple counterexample model \eqref{eq:4-7-04} which is covered by Theorem \ref{thm:4-04}.  Such counterexamples have their coefficients taking generic values, thus may introduce non-neglectable error to the field-counting method.

It is possible to incorporate Theorem \ref{thm:4-04} into the field-counting method.  But compared to $N_2$ and $N_0$ which are simple counts of fields with R-charge $2$ and $0$, identifying $P$'s and $Q$'s according to \eqref{eq:4-4-01} requires more elaboration of their appearance in superpotential terms.  Furthermore, counterexamples not covered by Theorem \ref{thm:4-04} exist~\cite{Brister:2022rrz}.  Therefore we stick to $N_2$ and $N_0$ to keep the field-counting method simple and efficient.

A straightforward way to check the accuracy of the field-counting method is to collect a large number of Wess-Zumino models with R-symmetries, examine their vacuum solutions, compare to the claims from the Nelson-Seiberg theorem and its extensions, and give the proportion of counterexamples in all the models.  Other information may also be extracted on request once the dataset of R-symmetric Wess-Zumino models is established.  Here we will present a preliminary construction of the dataset~\cite{Brister:2022vsz} in the folowing subsections.

\subsection{The data structure of R-symmetric Wess-Zumino models}

Perturbative, renormalizable and R-symmetric superpotentials are cubic polynomial functions of R-charge $2$.  The superpotential for fields $\{ \phi_i \mid i = 1, \dots , N \}$ with R-charges $\{ r_i \}$ can be expressed as
\begin{equation}
W = c + c_i \phi_i + c_{i j} \phi_i \phi_j + c_{i j k} \phi_i \phi_j \phi_k,
\end{equation}
where $c$, $c_i$, $c_{i j}$ and $c_{i j k}$ are generic coefficients for R-charge 2 terms, and zero for terms with other R-charges.  So an R-symmetric $W$ always has $c = 0$.  $W$ can be rewritten in a short notation
\begin{equation}
W = c_s \hat{\phi}^{\hat{a}^s}
  = c_s \phi_1^{a^s_1} \dots \phi_N^{a^s_N}, \label{eq:5-2-01}
\end{equation}
where $\hat{\phi} = \{ \phi_1, \dots , \phi_N \}$ is the ordered set of fields, or the array of fields, $\hat{a}^s = \{ a^s_1, \dotsc , a^s_N \}$ with $s = 1, \dots , m$ is the array of exponents in the $s$'th monomial term of $W$, and $m$ is the number of non-zero terms in $W$.  We only pick out terms with non-zero coefficients $c_s$ which respect the R-symmetry.  Renormalizability impose the constraint
\begin{equation}
a^s_1 + \dots + a^s_N \in \{ 0, 1, 2, 3 \}, \quad
a^s_i \in \{ 0, 1, 2, 3 \}, \label{eq:5-2-02}
\end{equation}
and the R-symmetry requires
\begin{equation}
a^s_1 r_1 + \dots + a^s_N r_N = 2. \label{eq:5-2-03}
\end{equation}
Because of the genericity requirement in parameters, we view superpotentials with the same polynomial form but different non-zero coefficients as the same model.  An R-symmetric Wess-Zumino model uniquely corresponds to a two-dimensional array of exponents $\hat{a} = \{ \hat{a}^s \} = \{ a^s_i \}$, and vise versa.  For now we focus on models without ambiguity in the R-charge assignment.  So each model uniquely corresponds to an array of R-charges $\hat{r} = \{ r_1, \dots , r_N \}$ which solves the R-charge equations \eqref{eq:5-2-03}.  And giving an array of R-charges $\hat{r}$, the array of exponents $\hat{a}$ can be uniquely determined by including all $\hat{a}^s$ which satisfy both \eqref{eq:5-2-02} and \eqref{eq:5-2-03}.  Therefore we can label R-symmetric Wess-Zumino models with $\hat{r}$.

Given the number of fields $N$, or the array of fields $\hat{\phi} = \{ \phi_1, \dots , \phi_N \}$, we collect all possible monomials up to cubic, or all possible array of exponents $\hat{A}^I = \{ A^I_1, \dots , A^I_N \}$ with $I = 1, \dots , M$, satifying
\begin{equation}
A^I_1 + \dots + A^I_N \in \{ 0, 1, 2, 3 \}, \quad
A^I_i \in \{ 0, 1, 2, 3 \}, \label{eq:5-2-04}
\end{equation}
and record them as a two-dimensional array of exponents $\hat{A} = \{ \hat{A}^I \} = \{ A^I_i \}$.  Each R-symmetric Wess-Zumino model corresponds to an array of indices $\hat{I} = \{ I_1, \dots , I_m \}$, which selects $m$ array of exponents $\hat{a} = \{ \hat{a}^s \} = \{ \hat{A}^{I_s} \}$ from $\hat{A}$ and safisfies
\begin{equation}
A^{I_s}_1 r_1 + \dots + A^{I_s}_N r_N = 2. \label{eq:5-2-05}
\end{equation}
Therefore models can be equivalently labeled with $\hat{I}$ given that $\hat{A}$ has been recorded.  Note that $m$, the length of $\hat{I}$ or the number of non-zero terms in $W$, can vary in different models even with the same number of fields $N$.

For simplicity, we use the array of R-charges $\hat{r}$ as the data format to represent R-symmetric Wess-Zumino models.  Each model is recorded as $\hat{r}$ being its uniquely fixed R-charge assignment.  The array of indices $\hat{I}$ for a model can be obtained from its recorded $\hat{r}$ if needed, by including all $I_s$ whose corresponding $\hat{A}^{I_s}$ satisfy \eqref{eq:5-2-05}.  To avoid redundancy among data records, R-charges $r_i$ in each array $\hat{r}$ are sorted so that R-charges $2$, $0$ and $- 2$ come first, followed by other R-charges in ascending order of their absolute values, with the positive R-charges preceding the negative R-charges if their absolute values are the same, i.e.:
\begin{equation}
\begin{split}
i < j \ &\Leftrightarrow \ (r_i = r_j)
                           \lor (r_i = 2)
                           \lor (r_i = 0 \land r_j \ne 2)
                           \lor (r_i = - 2 \land r_j \ne 2 \land r_j \ne 0)\\
        &\qquad
                           \lor (\lvert r_i \rvert < \lvert r_j \rvert \ne 2)
                           \lor (r_i = - r_j > 0). \label{eq:5-2-06}
\end{split}
\end{equation}
Indices $I_s$ in each array $\hat{I}$ are sorted in ascending order:
\begin{equation}
s < t \ \Leftrightarrow \ I_s < I_t. \label{eq:5-2-07}
\end{equation}
And arrays of exponents $\hat{A}^I$ in the two-dimensional array $\hat{A}$ are sorted in ascending order of their values represented with little-endian or the least significant digit first (LSDF) ordering, i.e.:
\begin{equation}
I < J \ \Leftrightarrow \ (A^I_k < A^J_k)
                          \land (A^I_i = A^J_i \ \text{for} \ i = k + 1, \dots , N). \label{eq:5-2-08}
\end{equation}

As an example, considering the $N = 2$ case, there are totally $M = 10$ possible monomials up to cubic:
\begin{equation}
\{ 1 \text{(a constant term)}, \phi_1, \phi_1^2, \phi_1^3, \phi_2, \phi_1 \phi_2, \phi_1^2 \phi_2, \phi_2^2, \phi_1 \phi_2^2, \phi_2^3 \},
\end{equation}
which are recorded as
\begin{equation}
\begin{split}
\hat{A} =& \{ \{ A^I_1, A^I_2 \} \mid I = 1, \dots , 10 \}\\
        =& \{ \{ 0, 0 \}, \{ 1, 0 \}, \{ 2, 0 \}, \{ 3, 0 \},
              \{ 0, 1 \}, \{ 1, 1 \}, \{ 2, 1 \},
              \{ 0, 2 \}, \{ 1, 2 \}, \{ 0, 3 \} \},
\end{split}
\end{equation}
The Witten's runaway model \eqref{eq:4-7-03} is recorded as
\begin{equation}
\hat{r} = \{ r_1, r_2 \}
        = \{ 2, -2 \},
\end{equation}
which gives
\begin{align}
\hat{I} &= \{ I_1, I_2 \}
         = \{ 2, 7 \},\\
\hat{a} &= \{ \hat{a}^1, \hat{a}^2 \}
         = \{ \hat{A}^{I_1}, \hat{A}^{I_2} \}
         = \{ \hat{A}^2, \hat{A}^7 \}
         = \{ \{ 1, 0 \}, \{ 2, 1 \} \}.
\end{align}
Thus the superpotential
\begin{equation}
W = c_1 \phi_1 + c_2 \phi_1^2 \phi_2
\end{equation}
can be reconstructed from the recorded data $\hat{r}$.  Redundant records from swapping fields such as
\begin{equation}
W = c_1 \phi_2 + c_2 \phi_1 \phi_2^2
\ \Leftrightarrow \ \hat{r} = \{ -2, 2 \}
\ \Leftrightarrow \ \hat{I} = \{ 5, 9 \}
\end{equation}
and from swapping terms in $W$ such as
\begin{equation}
W = c_1 \phi_1^2 \phi_2 + c_2 \phi_1
\ \Leftrightarrow \ \hat{r} = \{ 2, -2 \}
\ \Leftrightarrow \ \hat{I} = \{ 7, 2 \}
\end{equation}
has been eliminated by the ordering \eqref{eq:5-2-06}, \eqref{eq:5-2-07} and \eqref{eq:5-2-08}.

\subsection{Redundancy from reducible models}

Another type of redundancy comes from reducible models.  Such a model can be reduced to two or more subsystems which do not interact to each other, and each subsystem has been recorded as models with less fields.  A reducible model has its array of exponent $\hat{a}$ satisfying
\begin{equation}
\begin{gathered}
\hat{a} = \{ \hat{a}^1, \dots , \hat{a}^m \}
        = \{ \hat{a}^{s_1}, \dots , \hat{a}^{s_l} \}
          \cup \{ \hat{a}^{s_{l + 1}}, \dots , \hat{a}^{s_m} \},\\
a^{s_u}_i a^{s_{v}}_i = 0 \
\text{for} \
u \in \{ 1, \dots , l \}, \
v \in \{ l + 1, \dots , m \} \label{eq:5-3-01}
\end{gathered}
\end{equation}
for some $\left( \begin{smallmatrix} 1 & \dots & m\\ s_1 & \dots & s_m \end{smallmatrix} \right) \in S_m$ and $l \in \{ 1, \dots , m \}$.  This condition can be interpreted as the separation of $\hat{a}^s$ into two subsets orthogonal to each other.  Fields can also be separated into two groups with R-charges
\begin{equation}
\hat{r} = \hat{r}' \cup \hat{r}''
        = \{ r_{i_1}, \dots , r_{i_L}\}
          \cup \{ r_{i_{L + 1}}, \dots , r_{i_N} \}
\end{equation}
for some $\left( \begin{smallmatrix} 1 & \dots & N\\ i_1 & \dots & i_N \end{smallmatrix} \right) \in S_N$ and $L \in \{ 1, \dots , N \}$.  And with the rearranged indices, the array of exponent $\hat{a}$ satisfies
\begin{equation}
\begin{split}
a^{s_u}_{i_k} = 0 \
\text{for} \
&(u \in \{ 1, \dots , l \}, \ k \in \{ L+1, \dots , N \})\\
&\lor (u \in \{ l+1, \dots , m \}, \ k \in \{ 1, \dots , L \}).
\end{split}
\end{equation}
So we can write the reduction of $\hat{a}$ in the following notation:
\begin{equation}
\begin{gathered}
\hat{a} = \hat{a}' \oplus \hat{a}''
        = \{ \hat{a}'^1, \dots , \hat{a}'^l\}
          \oplus \{ \hat{a}''^1, \dots , \hat{a}''^{m - l} \},\\
\hat{a}'^u_k = \hat{a}^{s_u}_{i_k} \
\text{for} \
u \in \{ 1, \dots , l \}, \
k \in \{ 1, \dots , L \},\\
\hat{a}''^u_k = \hat{a}^{s_{l + u}}_{i_{L + k}} \
\text{for} \
u \in \{ 1, \dots , m - l \}, \
j \in \{ 1, \dots , N - L \}.
\end{gathered}
\end{equation}
For example, the model with four fields
\begin{equation}
\begin{gathered}
W = c_1 \phi_1 + c_2 \phi_2^3 + c_3 \phi_2 \phi_3 + c_4 \phi_1^2 \phi_4,\\
\hat{r} = \{ 2, 2 / 3, 4 / 3, - 2\}\\
\hat{a} = \{ \{ 1, 0, 0, 0 \}, \{ 0, 3, 0, 0 \},
             \{ 0, 1, 1, 0 \}, \{ 2, 0, 0, 1 \} \}
\end{gathered}
\end{equation}
has the reduction into two models:
\begin{equation}
\begin{gathered}
W = W' + W''
  = (c_1 \phi_1 + c_4 \phi_1^2 \phi_4) + (c_2 \phi_2^3 + c_3 \phi_2 \phi_3),\\
\hat{r} = \hat{r}' \cup \hat{r}''
        = \{ 2, - 2\} \cup \{ 2 / 3, 4 / 3\},\\
\hat{a} = \hat{a}' \oplus \hat{a}''
        = \{ \{ 1, 0 \}, \{ 2, 1 \} \} \oplus \{ \{ 3, 0 \}, \{ 1, 1 \} \}.
\end{gathered}
\end{equation}
If $\hat{r}'$ and $\hat{r}''$ has been recorded as models with two fields, recording $\hat{r}$ as a model with four fields becomes redundant, because it can be constructed from two decoupled sectors with $\hat{r}'$ and $\hat{r}''$.  To eliminate such redundancy, it is necessary and sufficient to ensure the recorded models never satisfy the condition \eqref{eq:5-3-01} for any rearrangement of indices.

\subsection{A brute-force search algorithm}

With the data structure specified in previous subsections, it is straightforward to construct the dataset with the following brute-force search algorithm:

\begin{enumerate}
\item Given $N$, Loop through integers $A^I_i$ satisfying the condition \eqref{eq:5-2-04}, and record all $M$ possibilities as a two-dimensional array $\hat{A} = \{ \hat{A}^I \} = \{ A^I_i \}$ with the ordering \eqref{eq:5-2-08};
\item Loop through combinations of $N$ indices $\hat{I} = \{ I_1, \dots, I_N \} \subset \{ 1, \dots, M \}$ which give $\hat{a} = \{ \hat{a}^s \} = \{ \hat{A}^{I_s} \}$ from the recorded $\hat{A}$, continue the following steps if the combination gives a unique solution $\hat{r}$ to the condition \eqref{eq:5-2-05};
      \begin{enumerate}
      \item Rearrange $\hat{r}$ with the ordering \eqref{eq:5-2-06};
      \item Loop through recorded models $\hat{r}^{(p)}$ and compare to $\hat{r}$, continue the next step if $\hat{r}$ is not recorded before;
      \item Loop through indices $I \in \{ 1, \dots, M \}$, append $\hat{A}^I$ to $\hat{a}$ and append $I$ to $\hat{I}$ if $\hat{A}^I$ satisfies \eqref{eq:5-2-05};
      \item Loop through rearrangements of indices $s$ in $\{ \hat{a}^s \}$ and check the condition \eqref{eq:5-3-01}, continue the next step if \eqref{eq:5-3-01} is never satisfied;
      \item Append $\hat{r}$ to the set of records $\{ \hat{r}^{(p)} \}$, rearrange $\hat{I}$ with the ordering \eqref{eq:5-2-07} and append it to to the set of records $\{ \hat{I}^{(p)} \}$;
      \end{enumerate}
\item Loop through record indices $p$;
      \begin{enumerate}
      \item Count $2$'s and $0$'s in $\hat{r}^{(p)}$ and identify the model as the $N_2 \le N_0$ type or the $N_2 > N_0$ type;
      \item Obtain $\hat{a} = \{ \hat{a}^s \} = \{ \hat{A}^{I^{(p)}_s} \}$ from $\hat{I}^{(p)}$, loop $h$ times to generate random coefficients $c_s$ for $W$ in \eqref{eq:5-2-01}, try to solve SUSY equations $\partial_i W = 0$ and identify the model as the SUSY type or the SUSY breaking type.
      \end{enumerate}
\end{enumerate}

In step 2, we have used the fact that $N$ linear equations with a full-rank coefficient matrix are needed to uniquely solve $N$ variables.  If $\hat{r}$ is uniquely fixed in a model, it must be possible to select $N$ monomial terms from $W$ with their exponents uniquely solve $\hat{r}$ by the equations \eqref{eq:5-2-03}.  So the exhaustion with all combinations of $N$ monomial indices gives all possible $\hat{r}$'s, and covers all R-symmetric Wess-Zumino models with $N$ fields.

For a given number of fields $N$, the number of monomials up to cubic is
\begin{equation}
M = \frac{1}{6} (N + 1) (N + 2) (N + 3)
  = O(N^3).
\end{equation}
It is also the time complexity of step 1 of the algorithm.  In step 2, the outermost loop is repeated to list all combinations of $N$ monomials from all the $M$ monomials obtained in the step 1.  The number of combinations equals the binomial coefficient
\begin{equation}
C = \binom{M}{N}
  = \frac{M!}{N! (M - N)!}
  = O(N^{2 N}),
\end{equation}
where the Stirling's approximation is used for the last equality.  All the inner loops in step 2 have polynomial time complexity with properly optimized algorithms.  So the total time complexity is $O(N^{2 N})$, and the algorithm belongs to the complexity class with exponential run time (EXP).  On the other hand, one can verify in polynomial run time whether a given $\hat{r}$ is in the dataset by constructing $\hat{a}$ from $\hat{r}$, checking whether $\hat{r}$ is the unique solution to \eqref{eq:5-2-05}, and checking the reducibility of $\hat{a}$.  Thus the problem belongs also to the complexity class with non-deterministic polynomial time (NP).

\subsection{Search result and interpretations}

In our preliminary study~\cite{Brister:2022vsz}, we constructed the dataset for $N \le 5$ using the brute-force search algorithm described in the previous subsection.  The search results is summarized in Table \ref{tb:5-01}.  The detailed list of models will be published in the follow-up work.

\begin{table}
    \centering
    \renewcommand\arraystretch{1.3}
    \begin{tabular}{|c|c|c|c|c|c|c|c|c|}
    \hline
    \multicolumn{3}{|c|}{number of fields}
    & $1$ & $2$ & $3$  & $4$   & $5$   & total\\
    \hline
    \multirow{4}{3.5em}{\centering number of models}
    & $N_2 \le N_0$              & SUSY
    & $2$ & $6$ & $19$ & $87$  & $474$ & $588$\\
    \cline{2-9}
    & \multirow{2}*{$N_2 > N_0$} & SUSY breaking
    & $1$ & $1$ & $7$  & $42$  & $266$ & $317$\\
    \cline{3-9}
    &                            & SUSY
    & $0$ & $0$ & $0$  & $1$   & $19$  & $20$\\
    \cline{2-9}
    & \multicolumn{2}{|c|}{total}
    & $3$ & $7$ & $26$ & $130$ & $759$ & $925$\\    
    \hline
    \end{tabular}
    \caption{Summary of search results.}
    \label{tb:5-01}
\end{table}

There are totally 925 R-symmetric Wess-Zumino models with $N \le 5$ fields.  For most models, field-counting gives correct predictions of SUSY or SUSY breaking, and the Nelson-Seiberg theorem and its extensions are verified.  About one third of models are of SUSY breaking type.  There are 20 counterexamples in the list including the simplest counterexample \eqref{eq:4-7-04} with $N = 4$.  All these counterexamples have $N_2 > N_0$, and give R-symmetry breaking SUSY vacua with $W = 0$ satisfying the bound of Theorem \ref{thm:4-05}.  So they are counterexamples to both the original Nelson-Seiberg theorem and the revised one.  Their R-charge assignments satisfy the condition in Theorem \ref{thm:4-04}.  The dataset for $N \le 5$ contains no counterexample beyond Theorem \ref{thm:4-04} like the one in~\cite{Brister:2022rrz}.

Since the counterexamples have SUSY vacua with $W = 0$, SUGRA extensions of these models give SUSY vacua with $V = 0$, and contribute to the the low-energy SUSY branch of the landscape.  It is proved that there is no counterexample of the other type which has $N_2 \le N_0$ and SUSY breaking~\cite{Li:2021ydn}.  So if we use the field-counting method through Theorem \ref{thm:4-03} to classify out SUSY vacua with a zero cosmological constant, all these counterexamples are false negatives, or type II errors in binary classification, which affect the recall but retain the precision of the classifier.  Furthermore, when SUGRA extensions of Wess-Zumino model are realized in Type IIB flux compactification theory, the R-symmetry breaking property of the vacua means that the expectation values of moduli deviate from the R-symmetric configuration of the Calabi-Yau manifold.  Then it is unnatural to turn on only R-symmetric fluxes and obtain an R-symmetric effective superpotential.  Therefore these counterexamples affect neither the recall nor the precision of the field counting method if we only consider R-symmetric SUSY vacua in the low-energy SUSY branch~\cite{Dine:2005gz, Dine:2004is, Dine:2005yq}, or string vacua with enhanced symmetries~\cite{DeWolfe:2004ns, DeWolfe:2005gy, Kanno:2017nub, Micu:2007rd, Palti:2007pm}.  Whether the R-symmetry breaking SUSY vacua have other string realizations is an open question.

\section{Outlooks}

In the previous section we have presented a preliminary dataset of R-symmetric Wess-Zumino models, with all R-charges uniquely fixed and the superpotentials being renormalizable.  We can construct a larger dataset if the renormalization requirement is relaxed.  For example, the quartic superpotential
\begin{equation}
W = a \phi_1 + b \phi_1^3 \phi_3 + c \phi_1 \phi_2 \phi_3
\end{equation}
uniquely fixes R-charges to 
\begin{equation}
\hat{r} = \{ r_1, r_2, r_3 \}
        = \{ 2, 4, -4 \},
\end{equation}
and all R-charge $2$ terms up to quartic has been included in $W$.  This model has $N_2 > N_0$ but gives a SUSY vacuum at
\begin{equation}
\phi_1 = 0, \quad
\phi_3 = - \frac{a}{c \phi_2}, \quad
\phi_2 \in \mathbb{C}, \quad
\text{with} \
V = W
  = 0.
\end{equation}
It is a counterexample to both Theorem \ref{thm:4-01} and Theorem \ref{thm:4-02} with only $N = 3$ fields.  It is not shown in Table \ref{tb:5-01} because of non-renormalizability.  If the superpotential is restricted to cubic to be renormalizable, counterexamples only begin to appear from $N = 4$ as shown in Table \ref{tb:5-01}.

Even within the scope of renormalizable models, our dataset still has some unexplored corners.  We have focused on models with all R-charges fixed, thus excluded models with ambiguity in the R-charge assignment, or equivalently models with an extra $U(1)$ symmetry, such as the example \eqref{eq:4-7-05}.  Whether the restriction from the extra symmetry introduces more counterexamples is worth studying.  Another issue is that we have only checked the SUSY equations $\partial_i W = 0$ to identify a model as SUSY or SUSY breaking type.  Some SUSY type models like the modified Polonyi model \eqref{eq:4-7-01} may have additional metastable SUSY breaking vacua.  There are also SUSY breaking type models like the Witten's runaway model \eqref{eq:4-7-03} which have only runaway directions but no metastable local minimum.  The good news is that if our purpose is to classify out models with SUSY vacua, both metastable SUSY breaking models and runaway models affect neither the recall nor the precision of the classifier.  These possibilities need to be considered if the dataset is used for other purposes.

Models with more fields requires more time for exhaustive search, and the run time grows exponentially.  With the time complexity $O(N^{2 N})$, the run time to exhaust models for $N \le 6$ and $N \le 7$ are respectively about $200$ and $70000$ times of the run time for $N \le 5$.  A better search algorithm and optimized code may reduce the run time, and the dataset with larger $N$ may be obtained in a reasonable time.  But unless there is an algorithm with polynomial time complexity, the brute-force search quickly become impracticable as $N$ goes larger, and a non-exhaustive sampling must be used to construct the dataset.  Finally, if SUGRA extensions of Wess-Zumino models are used as effective descriptions of Type IIB flux compactification in string phenomenology, the number of fields $N$ corresponds to the number of the complex structure moduli, or the Hodge number $h^{2, 1}$ of the Calabi-Yau manifold, which can reach several hundred for some known Calabi-Yau manifolds~\cite{CYdata:websites}.  It is challenging to make an unbiased sample of models with such a large $N$.  And we expect new methods such as machine learning~\cite{Ruehle:2020jrk} may help us to construct the dataset.

\section*{Acknowledgement}

The author thanks James Brister, Shihao Kou, Jinmian Li, Tianjun Li, Zhengyi Li, Xiao Liu, Xuewen Liu, Bo Ning, Dimitri Polyakov, Zipeng Tan, Greg Yang and Lu Yang for helpful discussions.  This work is supported by the National Natural Science Foundation of China under the grant number 11305110.

\end{document}